\documentclass[12pt]{elsarticle}                                          
\usepackage{graphicx}                                                         
\usepackage{a41}                                                                
\usepackage{xcolor}                     
\usepackage[rflt]{floatflt}
\usepackage{float}
\usepackage{lscape}
\usepackage{array}
\usepackage[english]{babel}
\usepackage[T1]{fontenc}
\usepackage{ae}
\usepackage{url}
\usepackage{amsmath, amsthm, amssymb}
\usepackage{slashed}

\usepackage{rotating}
\usepackage{graphicx}
\usepackage{comment}
\newcounter{mmacnt}
\def\restartmma{\setcounter{mmacnt}{0}}
\restartmma \catcode`|=\active
\def|#1|{\mathrm{#1}}
\catcode`|=12
\newenvironment{mma}{
\par\smallskip
\catcode`|=\active
\parskip=0pt\parindent=0pt 
\small
\def\In##1\\{%
\def\linebreak{\hfill\break\null\qquad}%
\refstepcounter{mmacnt}
\hangindent=2.5em\hangafter=0
\leavevmode
\llap{\tiny\sffamily In[\arabic{mmacnt}]:=\kern.5em}%
\mathversion{bold}\footnotesize$
\displaystyle##1$\normalsize
\mathversion{normal}\par
 }%
\def\Print##1\\{%
\def\linebreak{\hfill\break}%
\hangindent=2.5em\hangafter=0
\leavevmode ##1\par}%
\def\Out##1\\{%
\def\linebreak{$\hfill\break\null\hfill$}%
\kern\abovedisplayskip\par
\hangindent=2.5em\hangafter=0
\leavevmode
\llap{\tiny\sffamily Out[\arabic{mmacnt}]=\kern.5em}
\footnotesize$\displaystyle##1$
\normalsize\hfill\null\par
\kern\belowdisplayskip
}%
\def\Warning##1##2\\{%
\def\linebreak{\hfill\break}%
\hangindent=2.5em\hangafter=0
\leavevmode
{\scriptsize##1 : ##2}\par}%
}{%
\par\smallskip
}

\usepackage{color}
\newenvironment{fshaded}{%
\MakeFramed {\FrameRestore}
}%
{\endMakeFramed}

\makeatletter
\def\ps@pprintTitle{%
\let\@oddhead\@empty
\let\@evenhead\@empty
\def\@oddfoot{\reset@font\hfil\thepage\hfil}
\let\@evenfoot\@oddfoot
}
\makeatother
\usepackage{tikz}
\usetikzlibrary{matrix}
\allowdisplaybreaks[4]

\begin{document}
\begin{frontmatter}
\title{\Large
\textbf{
Neutral scalar pair productions 
through $W$-boson fusion at 
multi--TeV muon colliders
}}
\author[1,2]{Khiem Hong Phan}
\ead{phanhongkhiem@duytan.edu.vn}
\author[3]{Quang Hoang-Minh Pham}
\address[1]{\it Institute of Fundamental
and Applied Sciences, Duy Tan University,
Ho Chi Minh City $70000$, Vietnam}
\address[2]{Faculty of Natural Sciences,
Duy Tan University, Da Nang City $50000$,
Vietnam}
\address[3]
{\it VNUHCM-University of Science,
$227$ Nguyen Van Cu, District $5$,
Ho Chi Minh City $70000$, Vietnam}
\pagestyle{myheadings}
\markright{}
\begin{abstract} 
Full one-loop electroweak radiative corrections to
$\mu^- \mu^+ \to W^\pm W^\mp \to hh$ in the Standard Model
are computed for the first time in this work. We then
evaluate neutral scalar pair production through vector
boson fusion at multi--TeV muon colliders within
Two-Higgs-Doublet Model (THDM). In the phenomenological
analysis, the enhancement factor, defined as the ratio
of the cross sections for SM-like Higgs pair production
in the THDM with respect to the corresponding ones in
the SM, is examined over the viable regions of the
parameter space in the Type-X and Type-Y THDMs. Our
findings show that this ratio can reach a factor of $3$
in several regions within the valid parameter space of
the Type-X THDM, whereas it ranges from $0.91$ to $0.96$
over the entire parameter space of the Type-Y THDM.
Finally, we scan the cross sections for double CP-odd
Higgs production over the updated parameter space of
the Type-X and Type-Y THDMs. 
In the Type-Y case at $\sqrt{s} = 10$~TeV
with an integrated luminosity of
$\mathcal{L} = 10000~\text{fb}^{-1}$,
CP-odd Higgs pair production
in the $t\bar{t}b\bar{b}$ final state,
with subsequent top-quark decays into leptons
and bottom quarks taken into account,
can be tested with a statistical significance
exceeding the $2\sigma$ level at several
viable parameter points.
\end{abstract}
\end{frontmatter}
\section{Introduction}
Measurements of multi-scalar Higgs 
production are among the
main objectives at future colliders, including the
High-Luminosity Large Hadron Collider (HL-LHC),
future $e^{+}e^{-}$ lepton colliders,
and future multi--TeV muon colliders.
The measured data allows us to 
verify the structure of the scalar 
Higgs potential and 
offers deeper insight into
the electroweak symmetry-breaking
mechanism (EWSB) in particle physics. 
The ATLAS Collaboration has searched for 
SM-like Higgs boson pair production in the $b\bar{b}\gamma\gamma$ events at $\sqrt{s}=13~\text{TeV}$ 
in $pp$ collisions at the 
LHC~\cite{ATLAS:2021ifb,ATLAS:2023gzn}.
Additional measurements for SM-like Higgs boson 
pair production in association with a vector 
boson ($W$ or $Z$) at $\sqrt{s}=13~\text{TeV}$ 
have also been performed by the LHC~\cite{ATLAS:2022fpx}.
The CMS Collaboration has investigated nonresonant Higgs boson pair production in the four-bottom-quark final state in $pp$ collisions, as reported in Refs.~\cite{CMS:2022gjd,CMS:2022cpr}.
Further measurements of nonresonant Higgs boson pair production in the $b\bar{b}\tau^-\tau^+$~\cite{ATLAS:2022xzm} and four-bottom-quark~\cite{ATLAS:2023qzf} final states have been conducted by the ATLAS Collaboration.
A combined analysis of the $b\bar{b}\gamma\gamma$, $b\bar{b}\tau^-\tau^+$, and $b\bar{b}b\bar{b}$ channels has been presented in Ref.~\cite{ATLAS:2023vdy}.
The studies of Higgs pair production
through different final states have 
been also carried out
in Refs.~\cite{ATLAS:2023elc,
ATLAS:2024lsk, ATLAS:2024lhu,
ATLAS:2024pov,
CMS:2024fkb,
ATLAS:2025hhd, CMS:2024pjq}.

Phenomenological studies of Higgs boson 
pair production at the colliders within the SM 
and its extensions have been performed.
In particular, di-Higgs production in the 
THDM has been investigated at the LHC in 
parameter regions consistent with the 
diphoton excess and the muon $(g-2)$ 
anomaly~\cite{Iguro:2022fel}.  
Theoretical implications for Higgs 
boson pair production in the $b\bar{b}\mu^+\mu^-$ 
final state at the LHC have been discussed in
Ref.~\cite{Guo:2022biq}. 
Moreover, probing the electroweak 
sector using the process
$W_L W_L \to n \times h$ has been examined 
within various effective field theory (EFT) 
frameworks~\cite{Gomez-Ambrosio:2022qsi}.
Theoretical predictions for di-Higgs production
in final states with two $b$-tagged jets,
two leptons, and missing transverse momentum
at the HL-LHC have also been carried out
in Ref.~\cite{Huang:2022rne}.
Furthermore, double Higgs production via
$W$-boson fusion at TeV-scale
$e^+e^-$ colliders has been investigated within
effective field theory (EFT) frameworks,
including studies of sensitivity to
BSM Higgs couplings~\cite{Domenech:2022uud},
multi-Higgs measurements in the
SMEFT~\cite{Gomez-Ambrosio:2022why},
additional analyses of Higgs boson pair
production in EFTs~\cite{Alasfar:2023xpc},
as well as in extended Higgs sector
models~\cite{Abouabid:2021yvw}.  
Di-Higgs production has been studied in the Higgs singlet extension~\cite{Feuerstake:2024uxs} and the I(1+2)HDM Type-I~\cite{Arhrib:2025kml}. Further investigations in various BSM frameworks include vector-boson-fusion di-Higgs production in the SMEFT~\cite{Dedes:2025oda}, enhancements from TeV-scale heavy neutral leptons at future lepton colliders~\cite{Kriewald:2025bui}, and scenarios involving axion-like particles~\cite{Esser:2024pnc}. The potential of probing new physics through di-Higgs production at the LHC has also been explored in Ref.~\cite{Belfkir:2025jum}.

Higher-order corrections to Higgs boson pair production
are also of great interest in the literature.
In particular, QCD corrections
to Higgs boson pair production,
including decays into the
$b\bar{b}\tau^+\tau^-$ final state,
have been reported in Ref.~\cite{Li:2025gbx}.
In addition, double-logarithmic enhancements
in Higgs boson pair production in the high-energy
limit have been studied in Ref.~\cite{Hu:2025hfc}.
Next-to-leading-order electroweak corrections 
have been presented in 
Refs.~\cite{Bonetti:2025xtt, Braun:2025fpt},
while higher-order corrections to
Higgs boson pair production within the
Higgs Effective Field Theory (HEFT)
framework have been studied in
Refs.~\cite{Brivio:2025sib,Braun:2025hvr},
as well as top-Yukawa- and light-quark-induced
electroweak corrections in Ref.~\cite{Bhattacharya:2025egw}. 
Furthermore, NLO QCD and electroweak corrections to double Higgs production have been computed in 
Refs.~\cite{Bonciani:2018omm, Davies:2025ghl, Hu:2025aeo,
Jaskiewicz:2024xkd, Davies:2024znp, Bagnaschi:2023rbx, Davies:2021kex, Das:2020ujo, Wang:2020nnr, Baglio:2020wgt, Baglio:2020ini, Chen:2019fhs, Chen:2019lzz}.

It is clear that future lepton colliders (LCs)
could provide a cleaner environment than
hadron colliders, which are affected by large
QCD backgrounds. For this reason, future LCs
allow for higher-precision measurements.
Another aspect is that future multi--TeV 
muon colliders are proposed to access a 
higher-energy regime, enabling probes of 
new physics. Since scalar particles couple directly to vector bosons, vector boson fusion processes at multi--TeV muon colliders provide an excellent probe of additional scalar states predicted in many BSM scenarios, potentially enhancing the cross sections through $s$-channel resonance effects. For the above reasons, multi--TeV muon colliders could offer excellent prospects for measurements of double scalar production processes. So far, no calculations including 
full one-loop electroweak corrections for double scalar production via $W$-boson fusion have been reported. In this work, we evaluate for the first time full one-loop electroweak radiative corrections to the process 
$\mu^- \mu^+ \to W^\pm W^\mp \to hh$ in 
the Standard Model. Neutral scalar pair productions
through vector $W$-boson fusion
at multi--TeV muon colliders
within Two-Higgs-Doublet Models 
(THDM) are then computed and the 
corresponding cross sections 
are scanned over the updated 
parameter space of type-X and 
type-Y THDMs. The enhancement 
factor defined as the 
ratio of the cross sections
for SM-like Higgs pair productions
in the THDM with respect to the 
corresponding ones in the SM is
examined over the viable region of 
parameters in the types of the THDM.
Finally, CP-odd Higgs pair production
is studied over the parameter space 
of the models under investigation. 
In the Type-Y case at $\sqrt{s} = 10$~TeV
with an integrated luminosity of
$\mathcal{L} = 10000~\text{fb}^{-1}$,
CP-odd Higgs pair production
in the $t\bar{t}b\bar{b}$ final state,
with subsequent top-quark decays into leptons
and bottom quarks taken into account,
can be tested with a statistical significance
exceeding the $2\sigma$ level at several
viable parameter points.

The outline of this paper is as follows.
In Section~2, we present the THDM and 
discuss the relevant constraints.
Full one-loop electroweak radiative corrections
to the process $\mu^- \mu^+ \to W^\pm W^\mp \to hh$ 
in the SM are computed in detail in Section~3.
Phenomenological studies are presented in Section~4.
Finally, the conclusions and outlooks 
are given in Section~5. Checks of the 
calculations are reported in the appendices.
\section{The Two-Higgs-Doublet Model
and Its Constraints}
The structure of the models under investigation and the corresponding constraints are presented in this section. For further reviews of the THDM, including phenomenological studies, we refer the reader to Refs.~\cite{Tran:2025iur,Tran:2025zfq, Phan:2025pjt,Pham:2025ctj,Branco:2011iw}. In particular, the gauge and fermion contents of the THDM remain the same as in the SM. In contrast, the scalar Higgs sector is modified by introducing an additional scalar field with hypercharge $Y = 1/2$. With the addition of this scalar Higgs field, the most general scalar Higgs potential consistent with gauge invariance and renormalizability, and preserving a discrete
$Z_2$ symmetry, reads as follows 
(adopting the same notation as 
in Ref.~\cite{Phan:2025pjt}):
\begin{eqnarray} 
\label{potential}
\mathcal{V}
(\Phi_1, \Phi_2) &=&
\sum\limits_{j=1}^2
m^2_{jj}\Phi_j^\dagger\Phi_j
-
\left(m^2_{12}\Phi_1^\dagger\Phi_2
+
{\rm H.c.}\right)
+
\frac{1}{2}
\sum\limits_{j=1}^2
\lambda_j\left(\Phi_j^\dagger\Phi_j\right)^2
\nonumber \\
&&
\hspace{0cm}
+\lambda_3\Phi_1^\dagger\Phi_1\Phi_2^\dagger\Phi_2
+\lambda_4\Phi_1^\dagger\Phi_2\Phi_2^\dagger\Phi_1
+\left[\frac{1}{2}
\lambda_5\left(\Phi_1^\dagger\Phi_2\right)^2
+{\rm H.c.}\right].
\end{eqnarray}
As in our earlier works~\cite{Tran:2025iur, Tran:2025zfq, Phan:2025pjt, Pham:2025ctj}, the CP-conserving version of the THDM is employed for the phenomenological analysis. Consequently, all parameters, $m_{11}^2$, $m_{22}^2$, $m_{12}^2$, and $\lambda_1, \ldots, \lambda_5$, are taken to be real. In addition, a discrete $Z_2$ symmetry is introduced in the scalar potential, with only the allowed soft-breaking term, to explain the absence of tree-level flavor-changing neutral currents. As a result, four distinct types of THDMs are classified according to their different Yukawa coupling structures. Following Ref.~\cite{Tran:2025iur}, the general form of the Yukawa Lagrangian is taken as follows:
\begin{eqnarray}
\label{Yukawa}
-{\mathcal L}_\text{Y} 
&=&
\sum_{f=u,d,\ell}
\left(
\sum_{\phi_j=h, H}
\frac{m_f}{v}\xi_{\phi_j}^f
\phi_j {\overline f}f
-i\frac{m_f}{v}\xi_A^f
{\overline f}
\gamma_5fA
\right)
\\
&&
+
\frac{
\sqrt{2}
}{v}
\left[
\bar{u}_{i}
V_{ij}\left(
m_{u_i}
\xi^{u}_A P_L
+
\xi^{d}_A
m_{d_j} P_R \right)d_{j} H^+
\right]
\nonumber\\
&&
+ \frac{\sqrt{2}}{v}
\bar{\nu}_L
\xi^{\ell}_A
m_\ell \ell_R H^+
+ \textrm{H.c}.
\nonumber
\end{eqnarray}
Where $v$ denotes the vacuum expectation value.
The Cabibbo--Kobayashi--Maskawa (CKM) matrix, 
with elements $V_{ij}$, is taken into account 
in the above Lagrangian to describe quark mixing.
The operators $P_{L/R} = \frac{1 \mp \gamma_5}{2}$ correspond to projection operators, and the left- 
and right-handed leptons are denoted by $\ell_{L/R}$.
The coupling coefficients appearing in 
Eq.~\ref{Yukawa} are listed in Table~2 of 
Ref.~\cite{Aoki:2009ha}.

After the EWSB, additional scalar particles in the considered model include a CP-even Higgs ($H$), a CP-odd Higgs ($A$), and two singly charged Higgs bosons ($H^\pm$).
It is noted that the independent parameters in the THDM used in our analysis consist of the mixing angles $s_{\beta-\alpha}$ and $t_\beta$, the scalar masses $m_H$, $m_A$, and $m_{H^\pm}$, as well as the soft-breaking parameter $m_{12}^2$. Before proceeding to the study of neutral scalar pair production at multi--TeV muon colliders, we first provide an overview of the updated parameter space of the considered model. The six independent parameters of the THDM are constrained by both theoretical and experimental bounds. For a detailed discussion of these constraints and the corresponding results, we refer the reader to our previous works, Refs.~\cite{Tran:2025iur, Tran:2025zfq, Phan:2025pjt}. The parameters are scanned over the following ranges:
$m_{h}=125.09~\textrm{GeV}$,
$m_{H} \in [130, 1000]~\textrm{GeV}$,
$s_{\beta-\alpha} \in [0.97, 1]$, 
$m_{A,H^{\pm}} \in [130, 1000]~\textrm{GeV}$,
$t_{\beta} \in [0.5, 45]$, 
and $m_{12}^2 \in [0, 10^6]~\textrm{GeV}^2$.
The theoretical conditions are first 
imposed to scan the above parameter ranges 
with including perturbative 
unitarity, perturbativity, and vacuum stability 
of the model under investigation.
The valid data points are then tested 
against electroweak precision observables 
(EWPOs). In this test, the electroweak 
oblique parameters $S$, $T$, and 
$U$~\cite{Peskin:1991sw}, which are 
related to the $W$-boson 
mass~\cite{ParticleDataGroup:2024cfk}, 
are taken into account. 
The allowed parameter space is obtained 
by requiring the $S$, $T$, and $U$ 
parameters to bound within the $95\%$ 
confidence level limits.
It is worth mentioning that we 
employ {\tt 2HDMC-1.8.0}~\cite{Eriksson:2009ws} 
for all the aforementioned analyses.
Furthermore, to check that the theoretical predictions of the model under study are consistent with exclusion limits from Higgs searches at LEP, the LHC, and the Tevatron, we employ the public code {\tt HiggsBounds}-5.10.1~\cite{Bechtle:2020pkv}. This program matches theoretical predictions to experimental upper limits from collider searches established at the $95\%$ confidence level. Additionally, it is required that the theoretical predictions for the SM-like Higgs boson in the model be compatible with the corresponding experimental measurements at the LHC. To validate these constraints, we use the public code {\tt HiggsSignals}-2.6.1~\cite{Bechtle:2020uwn}.
Finally, experimental constraints from flavor physics are taken into account, in particular those arising from $B$-meson observables measured by LHCb. These constraints are using the {\tt SuperISO}~v4.1 package~\cite{Mahmoudi:2008tp}, and only parameter points whose theoretical predictions are compatible with the corresponding experimental data 
within the $2\sigma$ bounds retained.

\begin{figure}[H]
\centering
\begin{tabular}{cc}
\includegraphics[width=8.3cm,height=8cm]
{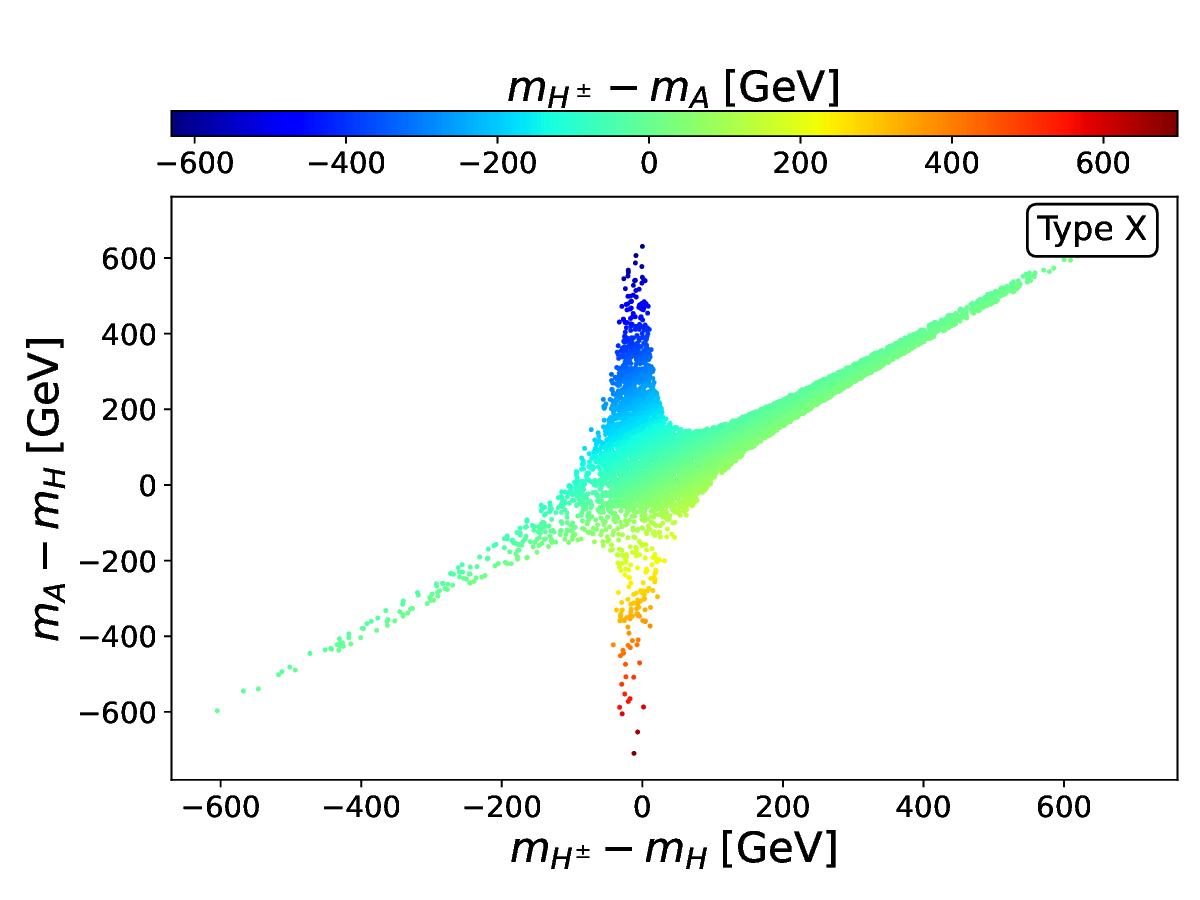}
&
\includegraphics[width=8.3cm,height=8cm]
{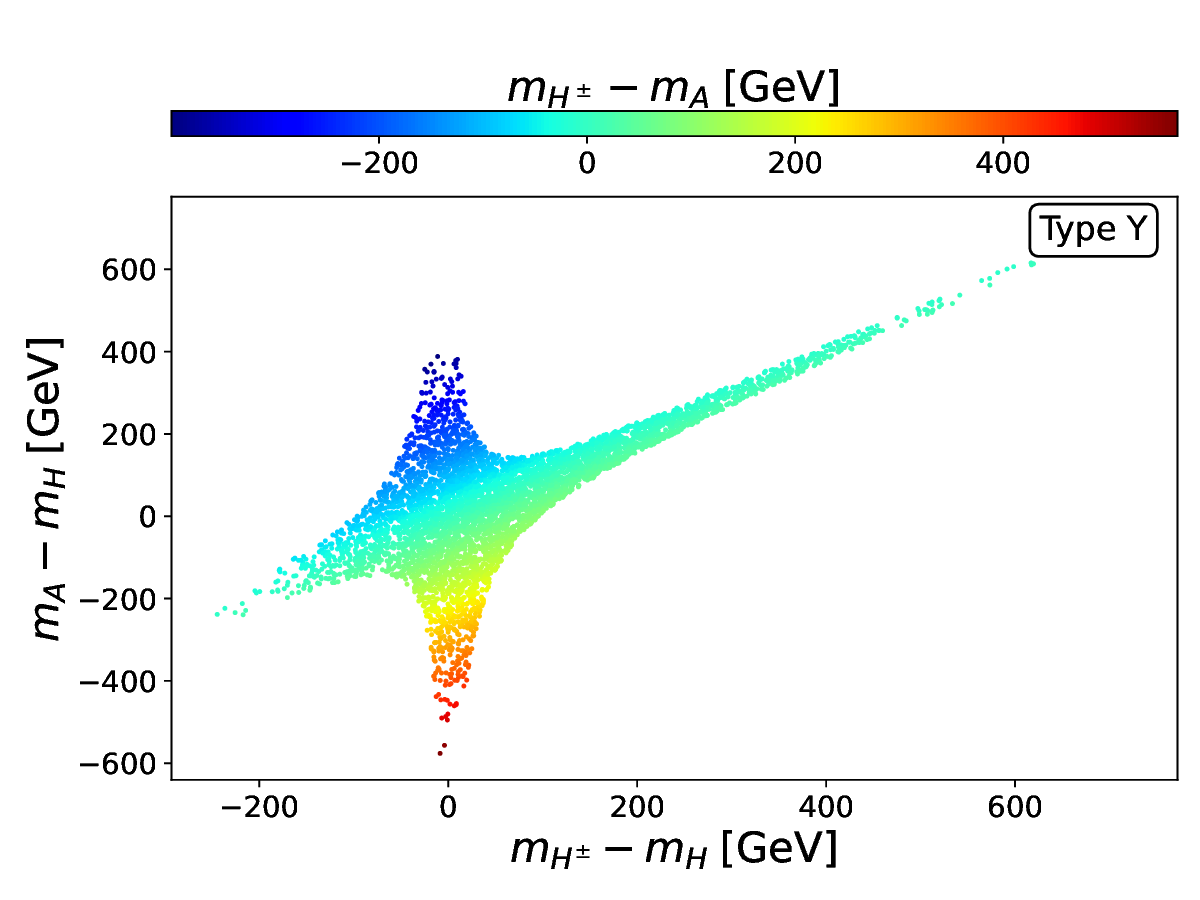}
\\
\includegraphics[width=8.3cm,height=8cm]
{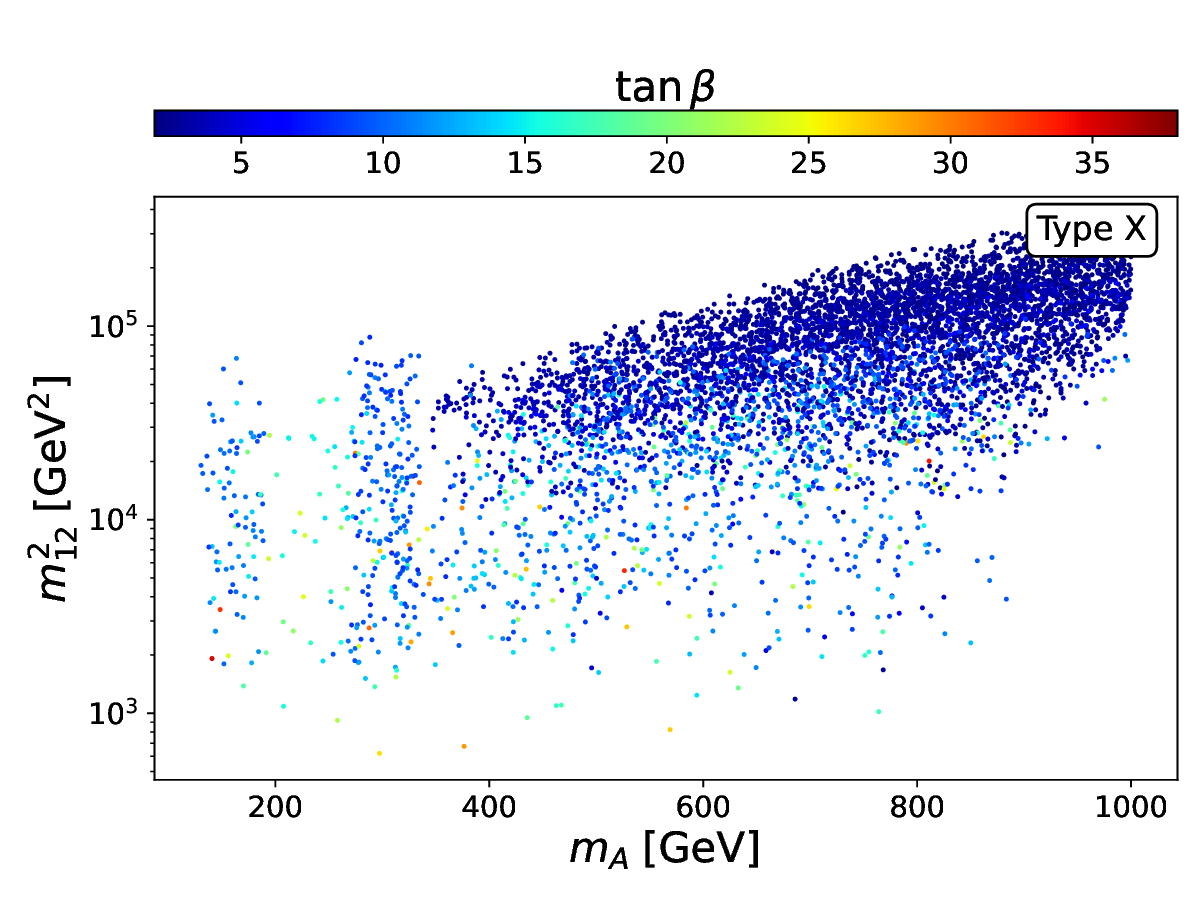}
&
\includegraphics[width=8.3cm,height=8cm]
{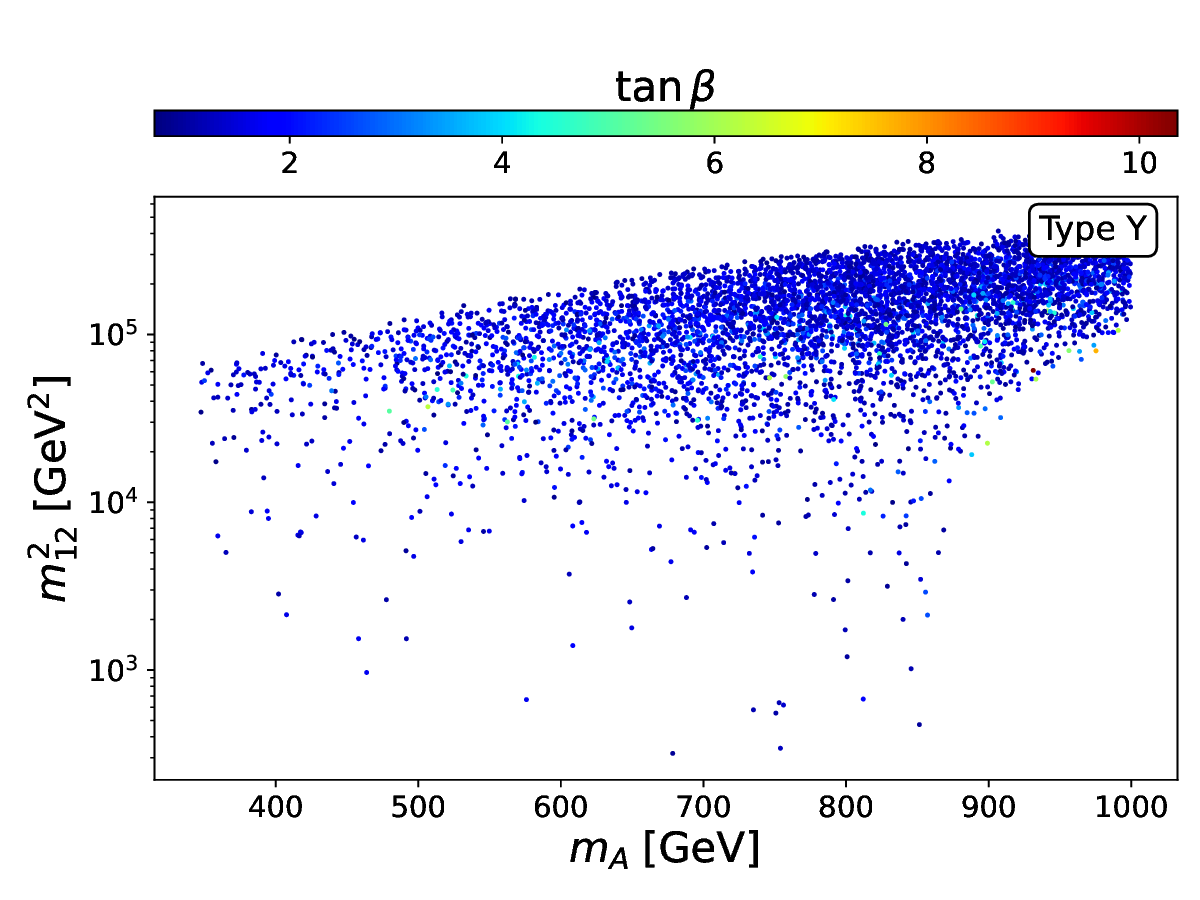}
\end{tabular}
\caption{
\label{scanplots}
The parameter space for six independent 
parameters of THDM type-X on the left 
and type-Y on the right. The upper-plots show
for scatter plots of $m_A-m_H, m_{H^\pm}-m_H$
and $m_{H^\pm}-m_A$. While lower-plots are for 
scatter plots of $m_{12}^2, m_A$ and 
the mixing angle $t_B$.  
}
\end{figure}

Using the updated parameter-space data files provided in
Refs.~\cite{Tran:2025iur, Tran:2025zfq, Phan:2025pjt},
we present several representative scatter plots that
are used in the subsequent analysis.
In Fig.~\ref{scanplots}, the parameter 
space of the six independent parameters
is shown for the Type-X THDM (left) and 
the Type-Y THDM (right). The upper panels
display scatter plots of
$m_A - m_H$, $m_{H^\pm} - m_H$, and $m_{H^\pm} - m_A$,
while the lower panels show scatter 
plots of $m_{12}^2$, $m_A$ and the mixing angle $\tan\beta$.
The results shown in the upper panels 
indicate that the mass splittings
among any two of the three scalar 
states are required to be smaller than
approximately $\sim 200$~GeV due to 
the EWPO constraints. In the Type-X
scenario, the mass splitting between 
the charged Higgs boson and the
CP-even Higgs boson can reach 
values as large as $\pm 600$~GeV.
Nevertheless, the dominant 
region in the $m_{H^\pm}-m_A$ plane is
concentrated in the green-shaded area, 
indicating that
$m_{H^\pm} \simeq m_A$ in this case. 
Similar features are observed for 
the mass splittings in the Type-Y THDM.
In this case, the allowed range of 
the mass difference is
$-600~\text{GeV} \leq m_A - m_H \leq 400~\text{GeV}$,
which is narrower than the corresponding range
$-600~\text{GeV} \leq m_A - m_H \leq 600~\text{GeV}$
found in the Type-X scenario.
In the lower panels, we present 
scatter plots of $m_{12}^2$, $m_A$, 
and $\tan\beta$ for the Type-X THDM 
(left) and the Type-Y THDM (right).
Over the full range of CP-odd Higgs
masses, the soft $Z_2$-breaking 
parameter $m_{12}^2$ lies in the range
$3\cdot 10^{3} \leq m_{12}^2 \leq 5\cdot 
10^{5}$ for both THDM types.
From these plots, we also find 
that $1 \leq \tan\beta \leq 20$.

\section{One-loop radiative corrections to
$\mu^- \mu^+ \to  W^{\pm} W^\mp
\to hh$ in SM}
In this section, we present the first calculations for 
full one-loop radiative corrections to 
$\mu^- \mu^+ \to W^\pm W^\mp \to hh$
within the SM in this work. The
calculations are performed with the help of the
{\tt GRACE-Loop} system, which is described in detail
in Ref.~\cite{Belanger:2003sd}. In general, the total
cross section for the process
$\mu^- \mu^+ \to VV \to hh$
can be computed as follows:
\begin{eqnarray}
\label{mastereq}
 \sigma_{\mu^- \mu^+ 
 \to VV \to hh}(s)
 &=&
 \sum\limits_{\lambda_1,\lambda_2}
 \int\limits_{4m_h^2/s}^1 
 d\tau
 \int\limits_{\tau}^1 
 \; 
 \frac{d\xi}{\xi}
 \;
 f_{V_{\lambda_1}/\mu}(\xi, Q^2)
 f_{V_{\lambda_2}/\mu}
 \left(
 \frac{\tau}{\xi}, 
 Q^2
 \right) 
 \hat{\sigma}
 _{V_{\lambda_1}V_{\lambda_2} 
 \to hh}(\hat{s}=\tau s).
 \nonumber\\
\end{eqnarray}
Where $\hat{\sigma}_{V_{\lambda_1}V_{\lambda_2} 
\to hh}(\tau s)$ 
is the corresponding cross section for
the partonic process $V_{\lambda_1}V_{\lambda_2} 
\to hh$. $\lambda_1$ and $\lambda_2$ are the polarization
degrees of the vector bosons.
In this work, the partonic processes 
are evaluated including one-loop radiative 
corrections using the {\tt GRACE-Loop}
system. Moreover, the vector-boson $V$ 
splitting functions describe the probability
for the emission of a weak vector boson from muon
which is socalled as the helicity-dependent 
parton distribution functions (PDFs)
are given by
\begin{eqnarray}
 f_{V/\mu}(\xi, Q^2)
 &=& \dfrac{g_V^2}{16\pi^2}
 \left[ \dfrac{1+(1-x)^2}{x}\ln\left(
 \frac{Q^2}{m_V^2} \right)
 +\dfrac{2 (1-x)}{x}
 \right]. 
\end{eqnarray}
Where $Q^2$ is energy scale. 
We emphasize that the first term corresponds
to the transverse polarization of 
vector boson PDF,
while the second term corresponds to the
longitudinal polarization
of the vector boson PDF. The coupling coefficient
$g_V$ is taken as $g_V^2 = \frac{g^2}{2}$ for
$V = W^\pm$, and $g_V^2 = \frac{g^2}{c_W^2}
\Big[(-1/2 + s_W^2)^2 + s_W^4/2\Big]$ for $V = Z$.
In general, several types of partonic subprocesses 
are considered in this work,
namely $\gamma\gamma \to hh$, 
$\gamma Z \to hh$, $ZZ \to hh$, and
$W^\pm W^\pm \to hh$. However, the first 
two channels are loop-induced
processes and are therefore expected 
to give smaller contributions
compared to the latter processes.
It is well known that the $ZZ$-fusion process proceeds as
$\mu^- \mu^+ \to ZZ \to \mu^- \mu^+ hh$, while the $WW$-fusion
channel is given by
$\mu^- \mu^+ \to W^{\pm} W^\mp \to \nu_{\mu}\bar{\nu}_{\mu} hh$.
These processes typically lead to different final states.
For the $ZZ$-fusion process, di-Higgs production is detected
in association with a muon pair, whereas for the latter fusion
process, the double Higgs boson signal is accompanied by
missing energy. From kinematic considerations, the
$ZZ$-fusion contribution is significantly smaller than the
$WW$-fusion contribution.
For these reasons, we focus on the calculation of the
partonic process $W^\pm W^\pm \to hh$. 
The relevant partonic processes are given by
\begin{eqnarray}
 \hat{\sigma}_{W^\pm W^\mp \to hh}(s) 
 &=& \int d \sigma_{W^\pm W^\mp \to hh}^{T}
 + \int d\sigma_{W^\pm W^\mp \to hh}^{V} 
 (\{\alpha, \beta, \cdots, \kappa\}, 
 C_{UV}, \lambda)
 \\
 &&
 + \int d \sigma_{W^\pm W^\mp \to hh}^{T}
 \delta_{\text{soft}}(\lambda\leq E_{\gamma_S}
 \leq k_c) 
 + \int d \sigma_{W^\pm W^\mp \to hh}^{H}(E_{\gamma_S}
 \geq k_c).
 \nonumber
\end{eqnarray}
In these formulas, $d\Phi_2$ denotes the $2 \to 2$
phase-space element. In the {\tt GRACE-Loop} program,
non-linear gauge-fixing terms have been implemented;
see Ref.~\cite{Belanger:2003sd} for more details.
As a result, the one-loop amplitude depends on the
non-linear gauge parameters and on the ultraviolet-divergent
parameter $C_{UV}$, whose dependence vanishes when all
contributions from the one-loop and counterterm diagrams
are taken into account. For the reaction
$W^\pm W^\mp \to hh$, virtual photon exchange occurs in the
loop, which leads to infrared divergences. The amplitude
depends on the photon mass regulator $\lambda$. By including
soft-photon radiation, the final results become independent
of $\lambda$. However, the results still depend on the
hard-photon cutoff $k_c$. To obtain the full one-loop
radiative corrections, hard-photon emission must also be
taken into account, corresponding to the channel
$W^\pm W^\mp \to hh\gamma$ with a hard photon in the final
state. The final results are free of all the
aforementioned parameters. For validation of the
calculation, we refer the reader to Appendix~A, where
explicit results are presented.

\begin{figure}[ht]
\centering
\begin{tabular}{c}
\includegraphics[width=11cm,height=6cm]
{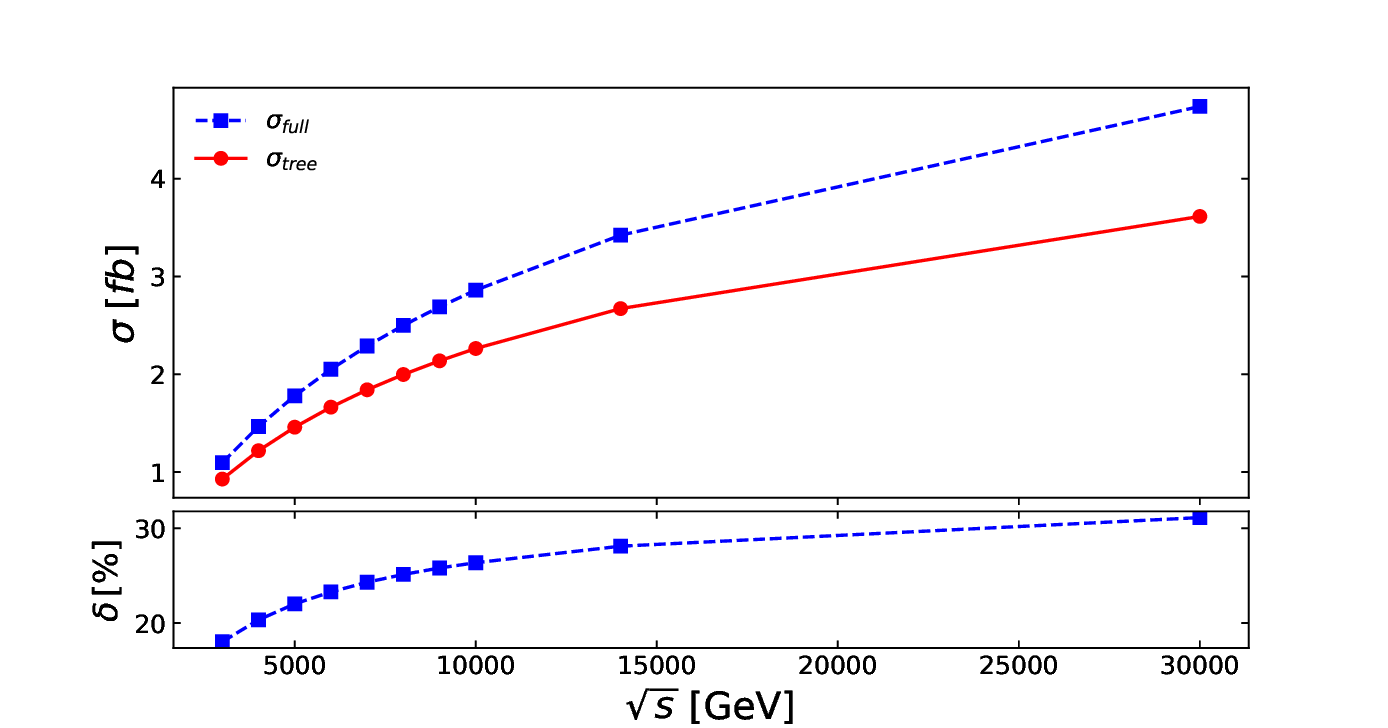}
\end{tabular}
\caption{ 
\label{cross-sections}
The cross sections for the 
process $\mu^- \mu^+ \to W^\pm W^\mp \to hh$ 
and full electroweak radiative corrections
are shown  as a function of the center-of-mass
(CoM) energy.}
\end{figure}

\begin{figure}[ht]
\centering
\begin{tabular}{cc}
\includegraphics[width=8.5cm,height=6cm]
{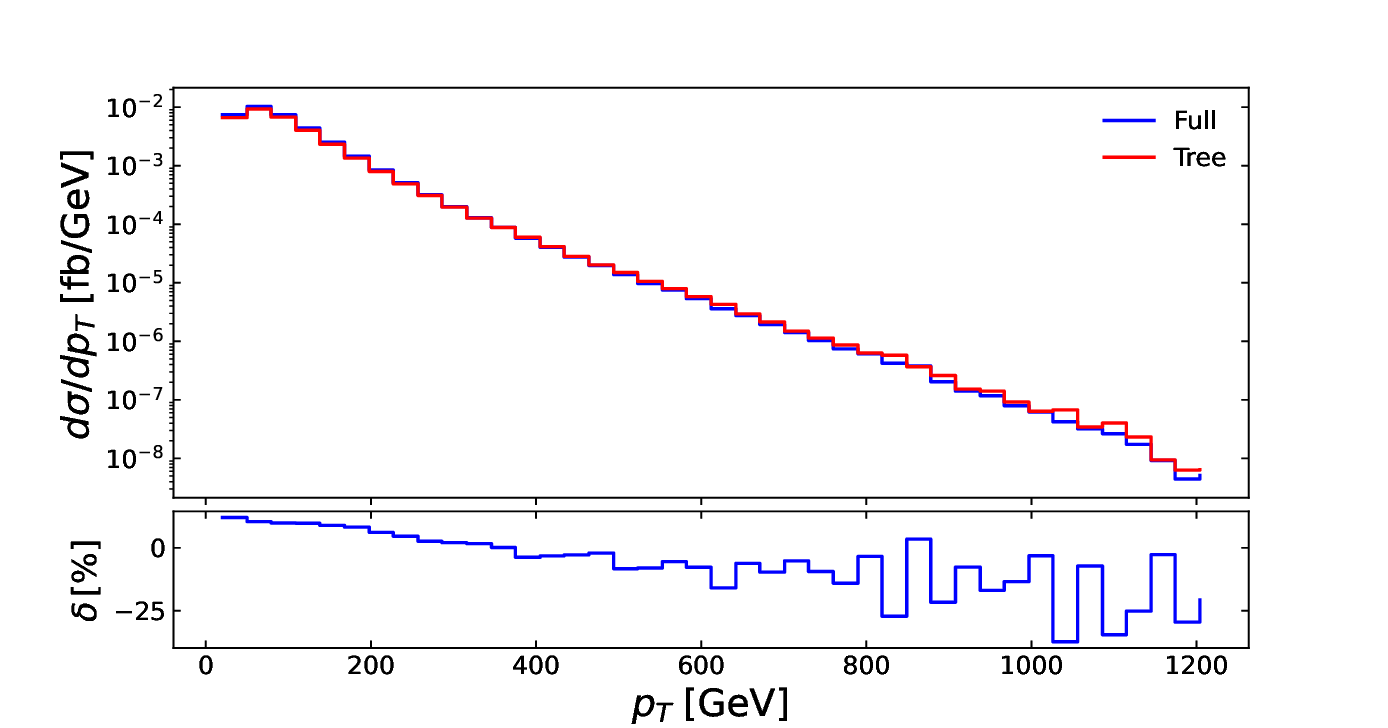}
&
\includegraphics[width=8.5cm,height=6cm]
{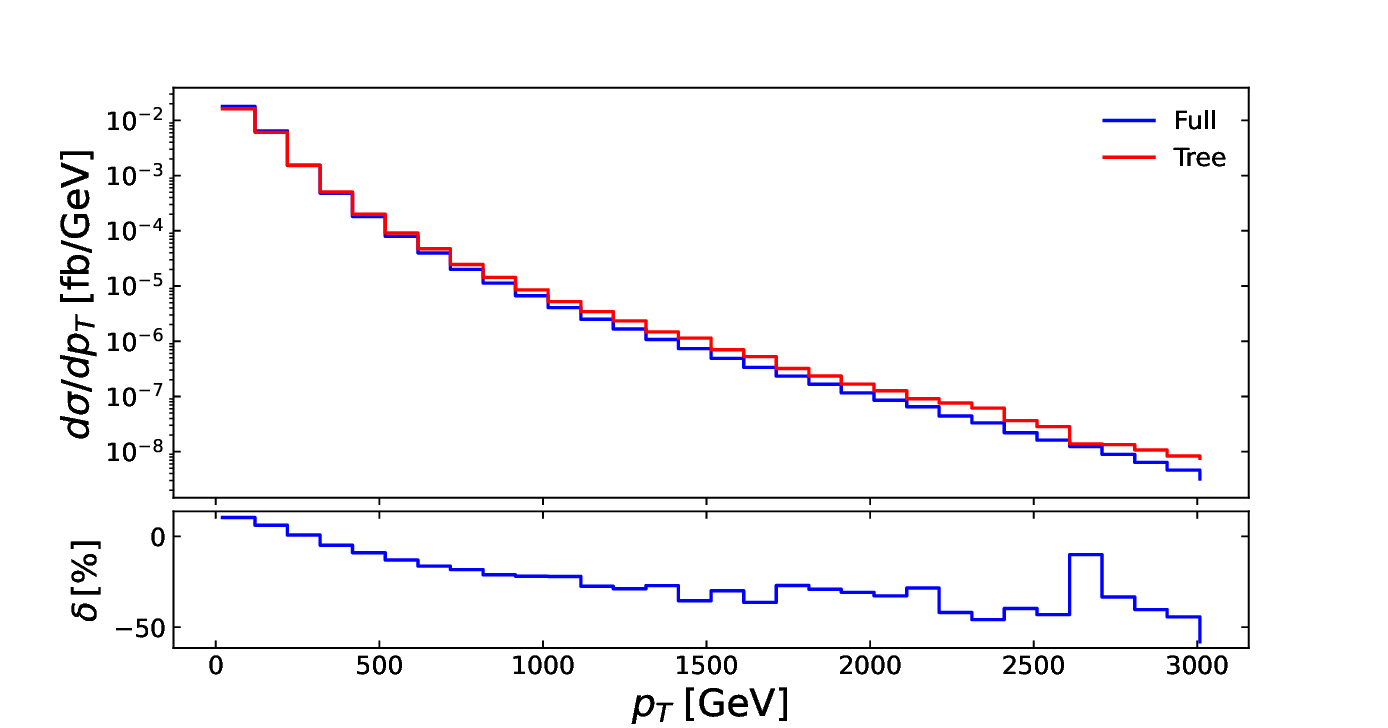}
\\
\includegraphics[width=8.5cm,height=6cm]
{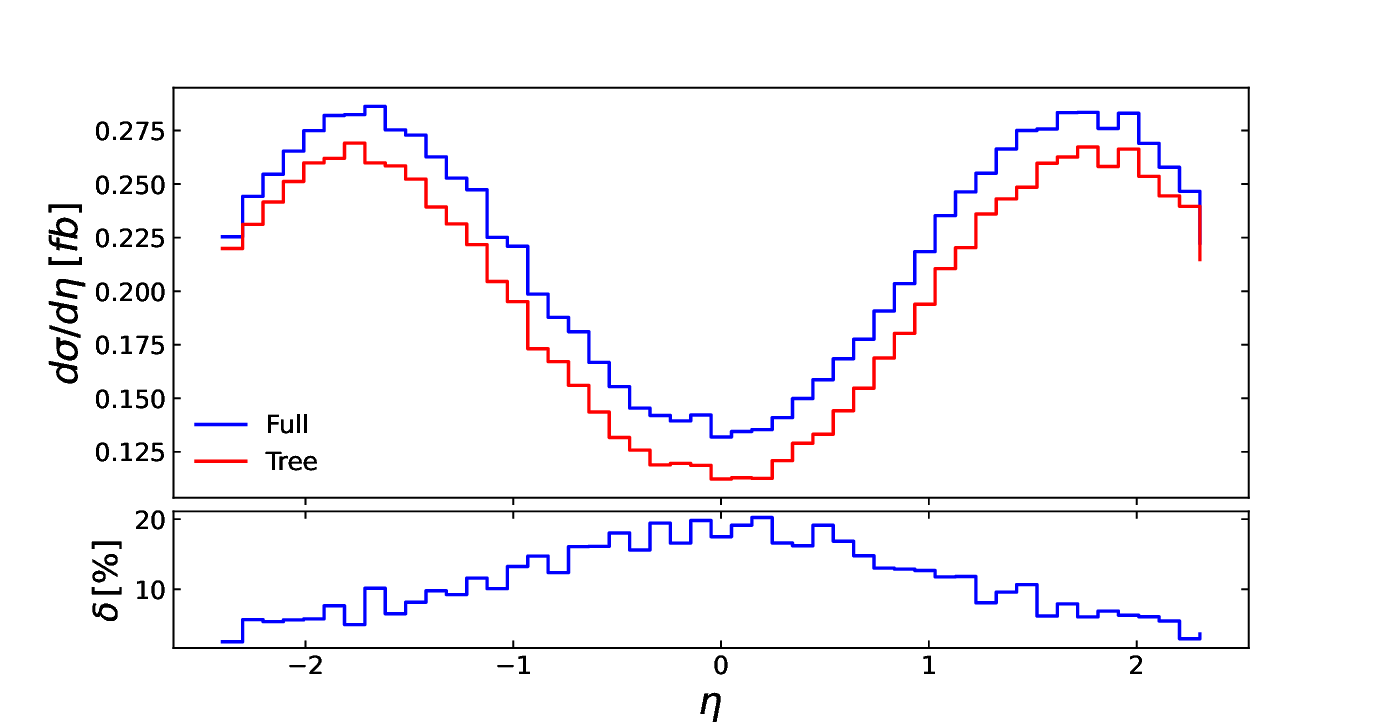}
&
\includegraphics[width=8.5cm,height=6cm]
{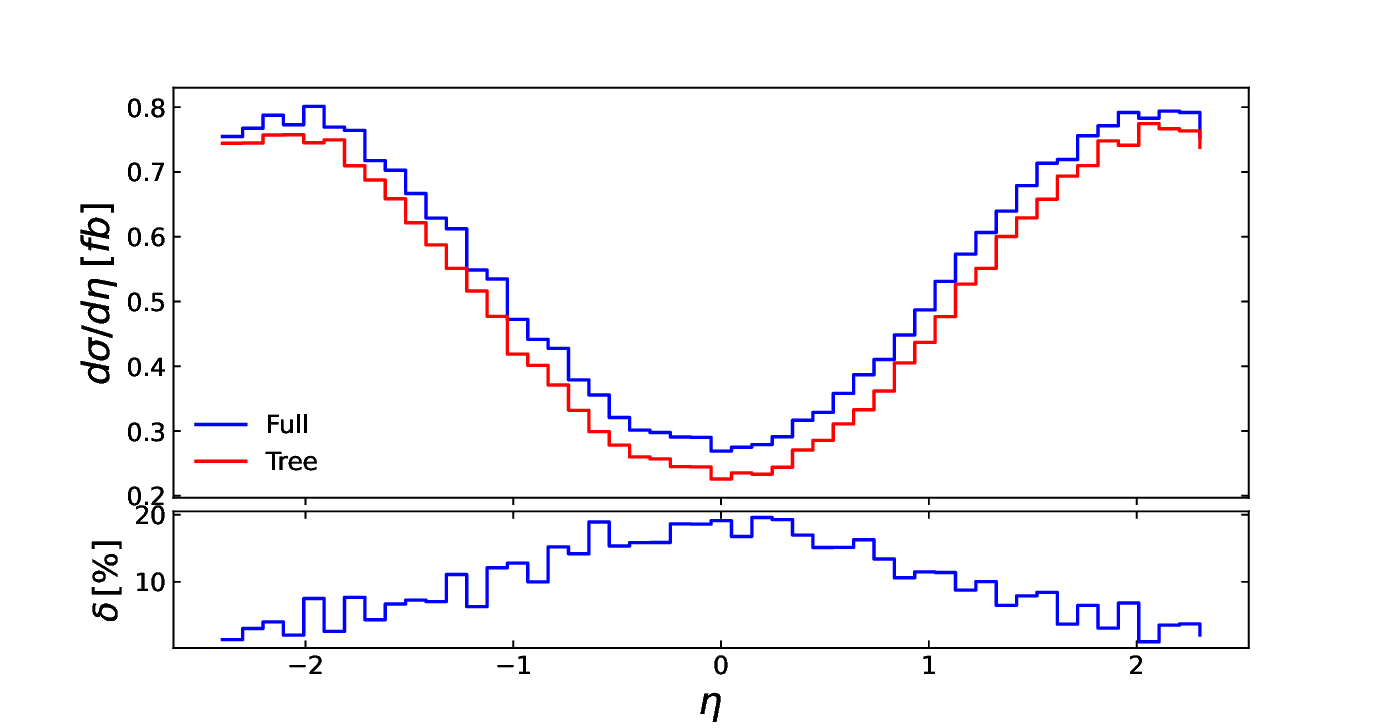}
\end{tabular}
\caption{\label{pTeta}
The differential cross sections with respect 
to $p_T^{h}$ and the rapidity $\eta^h$ are shown. 
The red line corresponds to the tree-level 
cross-sections, while the blue line represents 
the fully corrected cross sections.}
\end{figure}

Having the corrected partonic cross sections,
we evaluate the total cross section by convoluting 
them with vector W boson PDF as in Eq.~\ref{mastereq}.
We then investigate the effects of the full one-loop
electroweak radiative corrections to the processes
under consideration. In Fig.~\ref{cross-sections}, 
the cross sections for the
process $\mu^- \mu^+ \to W^\pm W^\mp \to hh$ are shown
as a function of the center-of-mass (CoM) energy.
In this Figure, the CoM energy is varied from
$3$~TeV to $30$~TeV. The red curve represents 
the tree-level cross section, while the blue curve 
shows the fully corrected cross section.
We find that the electroweak corrections range from
$20\%$ to $30\%$ as the CoM energy increases from
$3$~TeV to $30$~TeV.
The differential cross sections with respect 
to $p_T^{h}$ and the rapidity $\eta^h$
for SM-like Higgs are shown
in Figures~\ref{pTeta}. In these plots, the red 
line corresponds to the tree-level cross-sections, 
while the blue line represents the fully corrected 
cross sections. The left-panel plots show the 
distributions at $\sqrt{s}=3$ TeV, while the right-panel
plots present the corresponding ones at $\sqrt{s}=10$ TeV. 
Our finding is that the full electroweak corrections 
range from $\sim 10\%$ to $\sim -30\%$. In the high-$p_T$ 
tail region, the cross sections are very small,
and the corrections become meaningless in these 
regions. In principle, the electroweak corrections 
are of order $\sim 10\%$ for the distributions 
mentioned. We find similar corrections for the 
rapidity distributions.

We have checked that the longitudinal modes
$Z_L Z_L$ and $W^\pm_L W^\pm_L \to hh$
provide the dominant contributions in
the high-energy regime of future multi--TeV colliders.
This conclusion agrees with the numerical results in
Ref.~\cite{Ruiz:2021tdt}, in which
$W^\pm_L W^\pm_L \to hh$ provides a 
dominant contribution
of greater than $97\%$.
For this reason, the $Q^2$-dependence
of the cross section is rather small,
typically at the level of a few percent
(same results obtained 
in Ref.~\cite{Ruiz:2021tdt}). 
Last but not least, the cross sections presented in
Fig.~\ref{cross-sections} increase with 
developing CoM energies due to the Goldstone
equivalence theorem.
\section{Neutral scalar pair production through
vector boson fusion at multi--TeV muon colliders}
Neutral scalar pair production through
$W$-boson fusion at multi--TeV muon colliders
within the THDM framework is evaluated
in this section. In this work, we limit
our calculations to the tree-level cross 
section for neutral scalar pair production.
The amplitudes for the processes are generated by
using the {\tt FeynArts/FormCalc} 
packages~\cite{Hahn:2000kx,Hahn:1998yk}.
To verify our results, we generate the amplitudes
in the general $R_\xi$ gauge. The results are then tested 
for gauge invariance by varying 
the gauge-fixing parameter $\xi$. 
After successfully generating the corresponding 
amplitudes, we proceed to analyze the phenomenological 
results in the following subsections.
\subsection{$\mu^- \mu^+ \to W^{\pm} W^\mp \to hh$}%
We first consider SM-like Higgs pair production
via $W$-boson fusion at multi--TeV muon colliders.
For both signal and background processes, the
total cross sections for
$\mu^- \mu^+ \to W^\pm W^\mp \to hh$ are given by
Eq.~\ref{mastereq}. In the phenomenological
analysis, we study the enhancement factor,
which is defined as follows:
\begin{eqnarray}
\label{enhanceFactor}
 \mu_{hh}
 &=& 
 \dfrac{\sigma_{\mu^- \mu^+ 
 \to W^{\pm} W^\mp \to hh }^{THDM}
 }{\sigma_{\mu^- \mu^+ 
 \to W^{\pm} W^\mp \to hh }^{SM}}
 (\mathcal{P}_{\textrm{THDM} }). 
\end{eqnarray}
Where $\mathcal{P}_{\textrm{THDM}} = 
\{s_{\beta-\alpha}, t_\beta, m_H, m_A, 
m_{H^\pm}, m_{12}^2\}.$
We are going to study the enhancement 
factor in the viable parameter space of 
the THDM obatined in above section.
In Fig.~\ref{enhanceFactor}, 
the enhancement factor
$\mu^{hh}$ is scanned over 
$\tan\beta$ and $m_{H}$ 
for the Type-X THDM shown in 
the left panels and for 
the Type-Y THDM 
shown in the right panels.
The results are presented 
at a center-of-mass energy
of $3$~TeV in the two upper
plots and at $10$~TeV
in the two lower plots.
It is noted that we apply 
$p_T^{h}\geq 20$ GeV and 
peudo rapidity $|\eta_{h}|<2.4$.

The results indicate that
the enhancement factor
can reach values of up 
to about $3$ for several 
viable parameter points 
in the Type-X THDM, whereas 
it is always smaller than 
unity in the Type-Y THDM.
For the Type-X THDM, we 
find that the enhancement factor
tends to be close to unity 
when $\tan\beta \leq 5$
and for $m_H \geq 400$~GeV,
whereas it lies in the range
$0.91$--$0.95$ for the Type-Y THDM.
In general, this process is the 
same for the Type-X and Type-Y THDMs, 
since it does not involve scalar--fermion
couplings. The main difference 
in the observed behavior arises from
the distinct allowed parameter spaces, 
which are primarily constrained by 
the Yukawa couplings of 
the scalars to fermions.
These results offer interesting prospects for
distinguishing between the two types of THDMs
through measurements of the production processes
at future multi--TeV colliders.
\begin{figure}[ht]
\centering
\begin{tabular}{cc}
\includegraphics[width=8.5cm,height=6cm]
{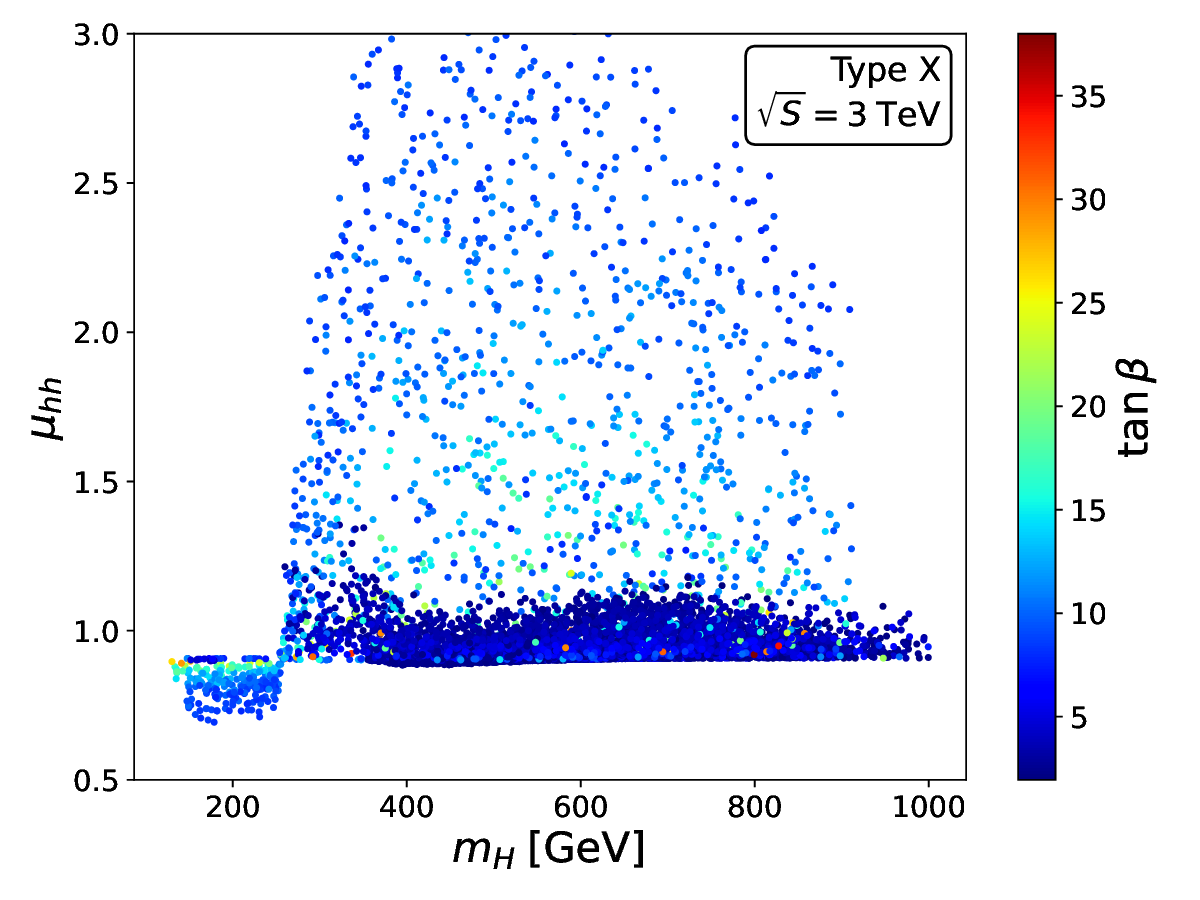}
&
\includegraphics[width=8.5cm,height=6cm]
{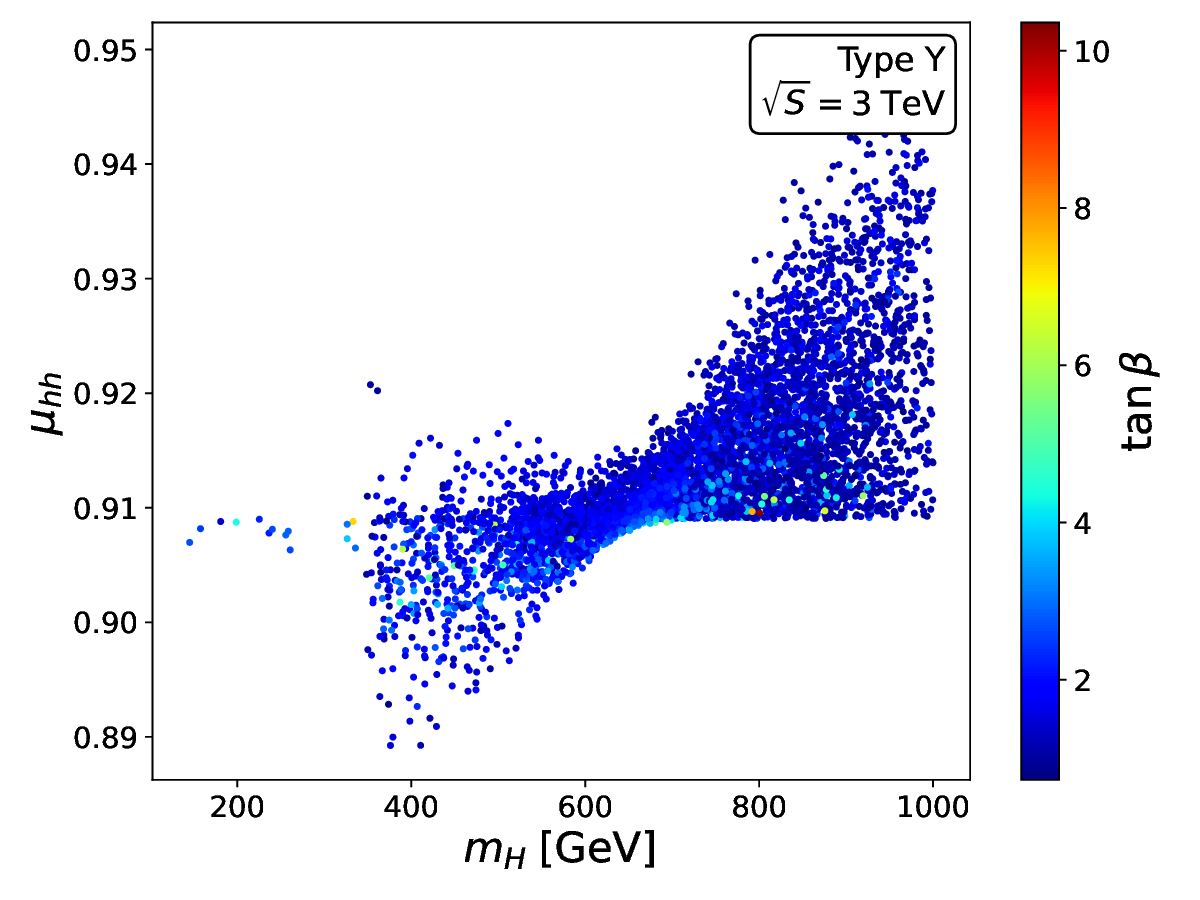}\\
\includegraphics[width=8.5cm,height=6cm]
{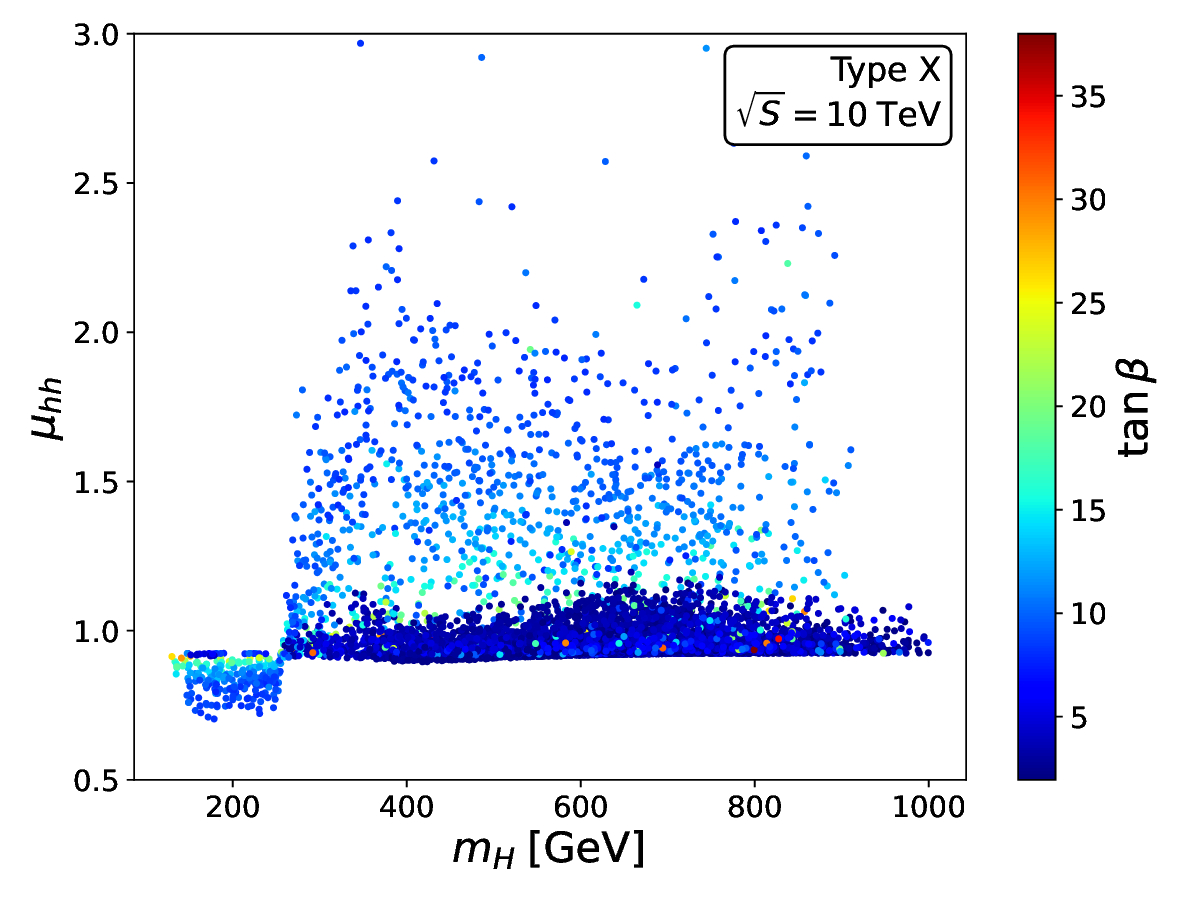}
&
\includegraphics[width=8.5cm,height=6cm]
{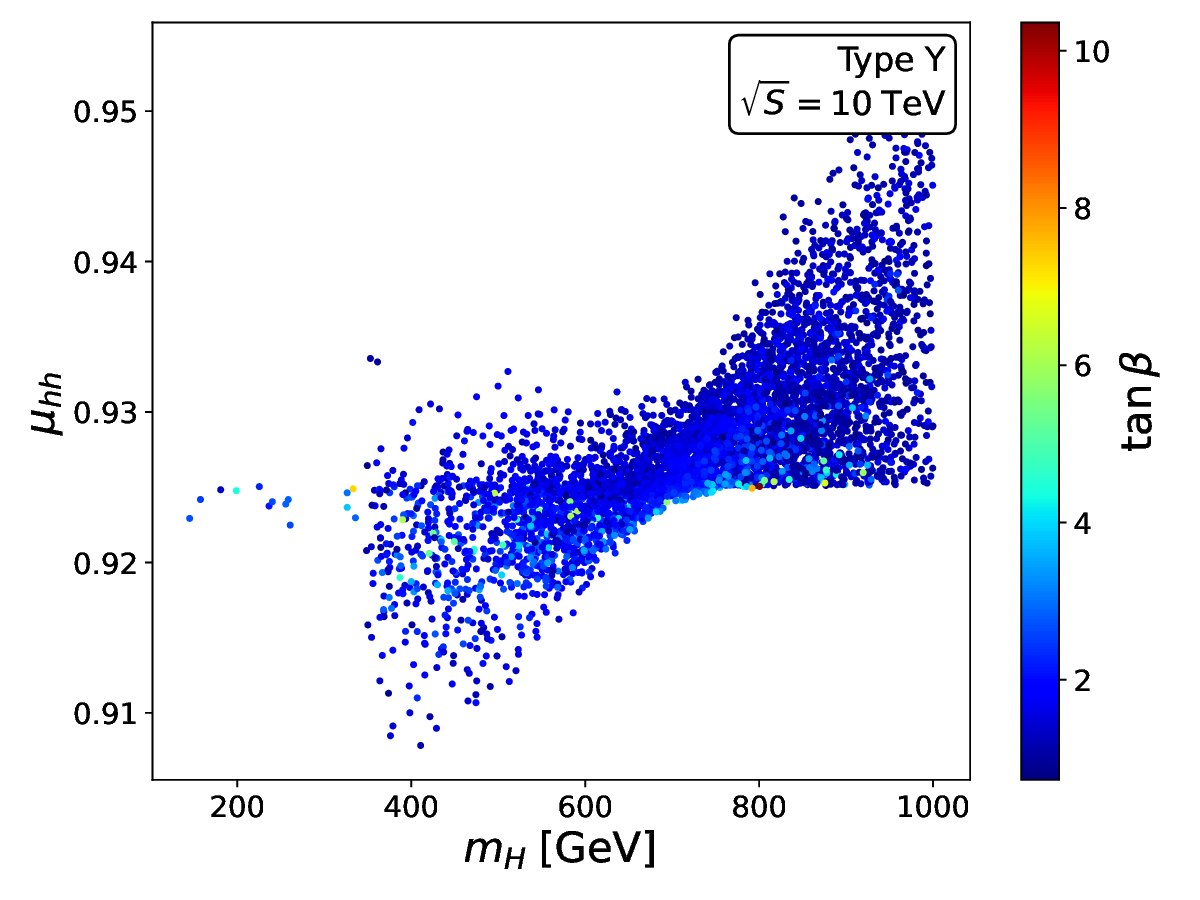}
\end{tabular}
\caption{ \label{enhanceFactor}
The enhancement factor $\mu_{hh}$ is 
scanned over the $\tan\beta$ and $m_{H}$ 
parameter space,
with the Type-X THDM shown in the left panels
and the Type-Y THDM shown in the right panels.
The results are presented at a center-of-mass energy
of $3$~TeV in the two upper plots and at $10$~TeV
in the two lower plots.
}
\end{figure}
\subsection{$\mu^- \mu^+ \to W^\pm W^\mp
\to AA \to t\bar{t}b \bar{b}$} 
In this subsection, we present the computation of
the cross sections for
$\mu^- \mu^+ \to W^\pm W^\mp \to AA \to t\bar{t} b\bar{b}$.
As demonstrated in the following paragraphs, we
consider the decay channels of the CP-odd Higgs,
$A \to t\bar{t}$ and $A \to b\bar{b}$. The cross
section for
$\mu^- \mu^+ \to W^\pm W^\mp \to A(p_3)A(p_4) \to
t\bar{t} b\bar{b}$ is evaluated as follows:
\begin{eqnarray}
\label{master}
\sigma_{t\bar{t}b \bar{b}}(s)
&=&
\sum\limits_{\lambda_1,\lambda_2}
 \int\limits_{\frac{4m_A^2}{s}}^1 
 d\tau
 \int\limits_{\tau}^1 
 \; 
 \frac{d\xi}{\xi}
 \;
 f_{W_{\lambda_1}/\mu}(\xi, Q^2)
 f_{W_{\lambda_2}/\mu}
 \Big(\frac{\tau}{\xi}, Q^2
\Big) \times
\nonumber\\
&&
\times (2!)
\int\limits_{4m_t^2}^{\hat{s}}
\frac{dp_3^2}{\pi}
\dfrac{m_A \Gamma_{A\to t\bar{t}}}
{(p_3^2-m_A^2)^2+\Gamma^2_{A} m_A^2}
\times 
\nonumber\\
&&
\times 
\int\limits_{4m_b^2}^{(\sqrt{\hat{s}}-
\sqrt{p_3^2})^2}
\frac{dp_4^2}{\pi}
\dfrac{m_A \Gamma_{A\to b\bar{b}}}
{(p_4^2-m_A^2)^2+\Gamma^2_{A} m_A^2}
\times 
\nonumber\\
&&
\times 
\frac{1}{2!}\int d\Phi_3
\Big|
\mathcal{M}_{W_{\lambda_1}
W_{\lambda_2} 
\to 
A(p_3)A(p_4)\to 
t\bar{t}b \bar{b}
}(\hat{s}=\tau s)
\Big|^2.
\end{eqnarray}
Here, $\Gamma_{A\to b\bar{b}}$, $\Gamma_{A\to t\bar{t}}$,
and $\Gamma_{A}$ denote the partial decay widths of the
CP-odd Higgs boson into $b\bar{b}$ and $t\bar{t}$, and
its total decay width, respectively. All the decay widths
mentioned above are calculated using the public code
{\tt H-COUP}~\cite{Aiko:2023xui}. Note that a factor of $2$
is included to account for the interchange of the decay
modes $A \to b\bar{b}$ and $A \to t\bar{t}$, while an
additional factor of $1/2$ arises from the presence of two
identical CP-odd Higgs bosons in the final state of the
partonic process $W^\pm W^\mp \to AA$.
We also emphasize that off-shell CP-odd Higgs bosons
(i.e., $p_3^2 \neq m_A^2$ and $p_4^2 \neq m_A^2$) are taken
into account in this procedure. The partonic process is
first generated in the general $\mathcal{R}_{\xi}$ gauge
with the help of the computer algebra packages
{\tt FeynArts/FormCalc}~\cite{Hahn:2000kx,Hahn:1998yk}.
The results are then checked for consistency by varying
the $\xi$-gauge parameters. The corrected amplitude for
the process under consideration is convoluted as in
Eq.~\ref{master}.

\begin{figure}[ht]
\centering
\begin{tabular}{cc}
\includegraphics[width=8.5cm,height=6cm]
{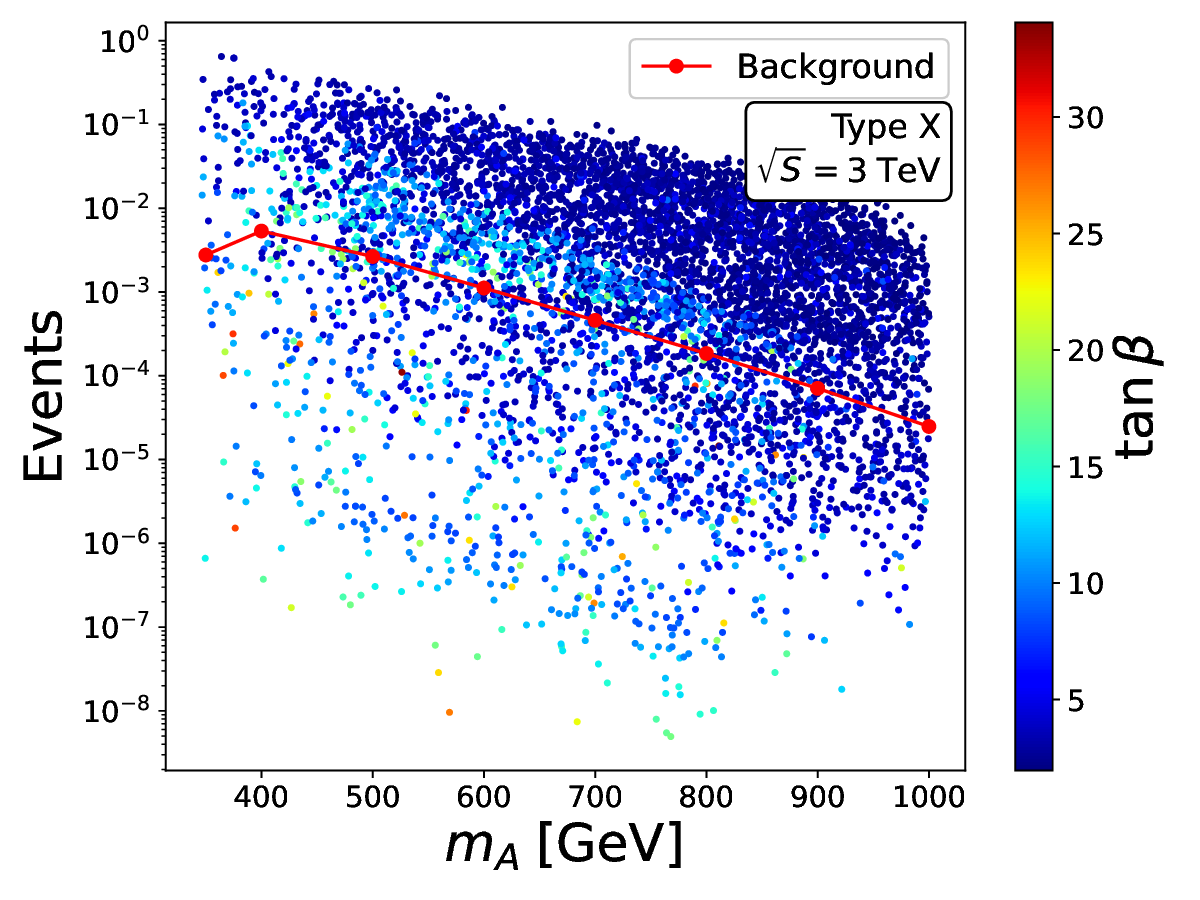}
&
\includegraphics[width=8.5cm,height=6cm]
{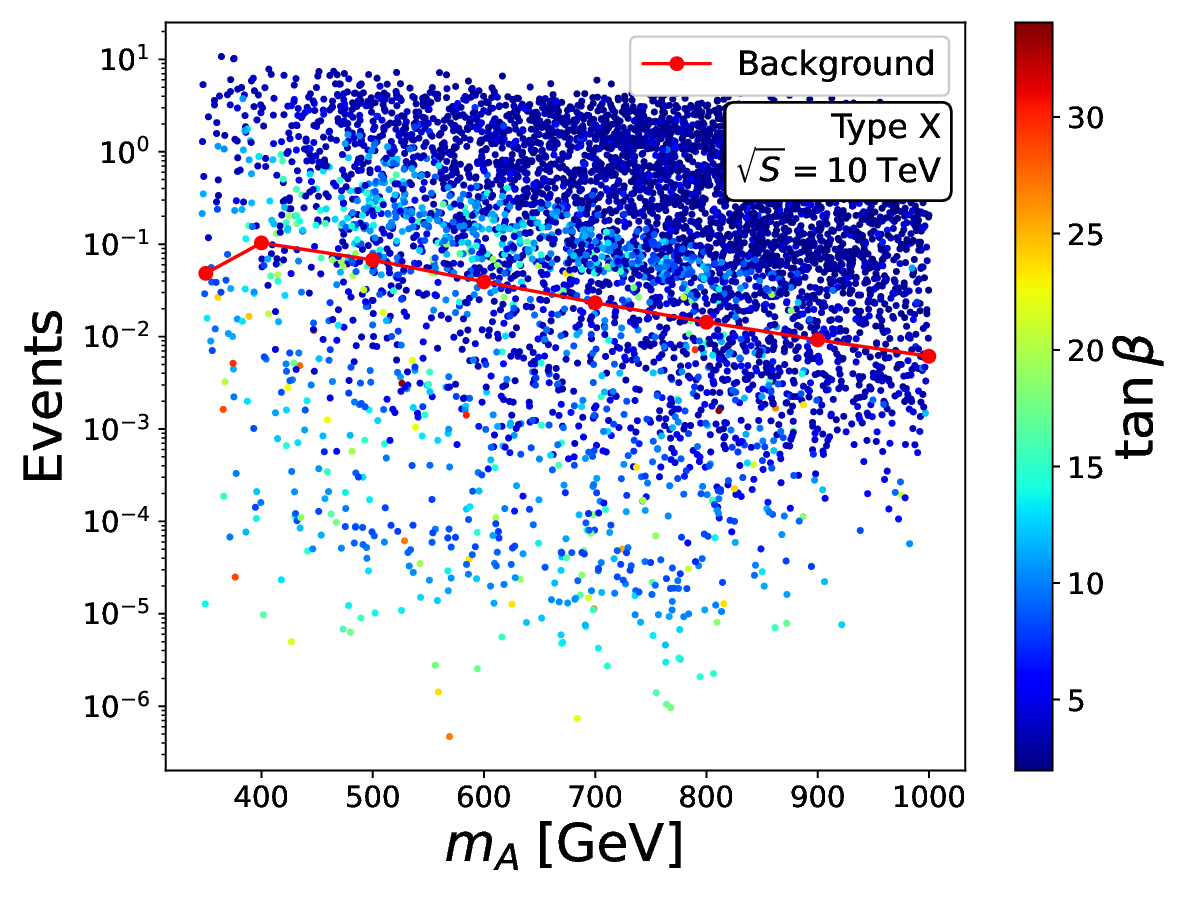}
\\
\includegraphics[width=8.5cm,height=6cm]
{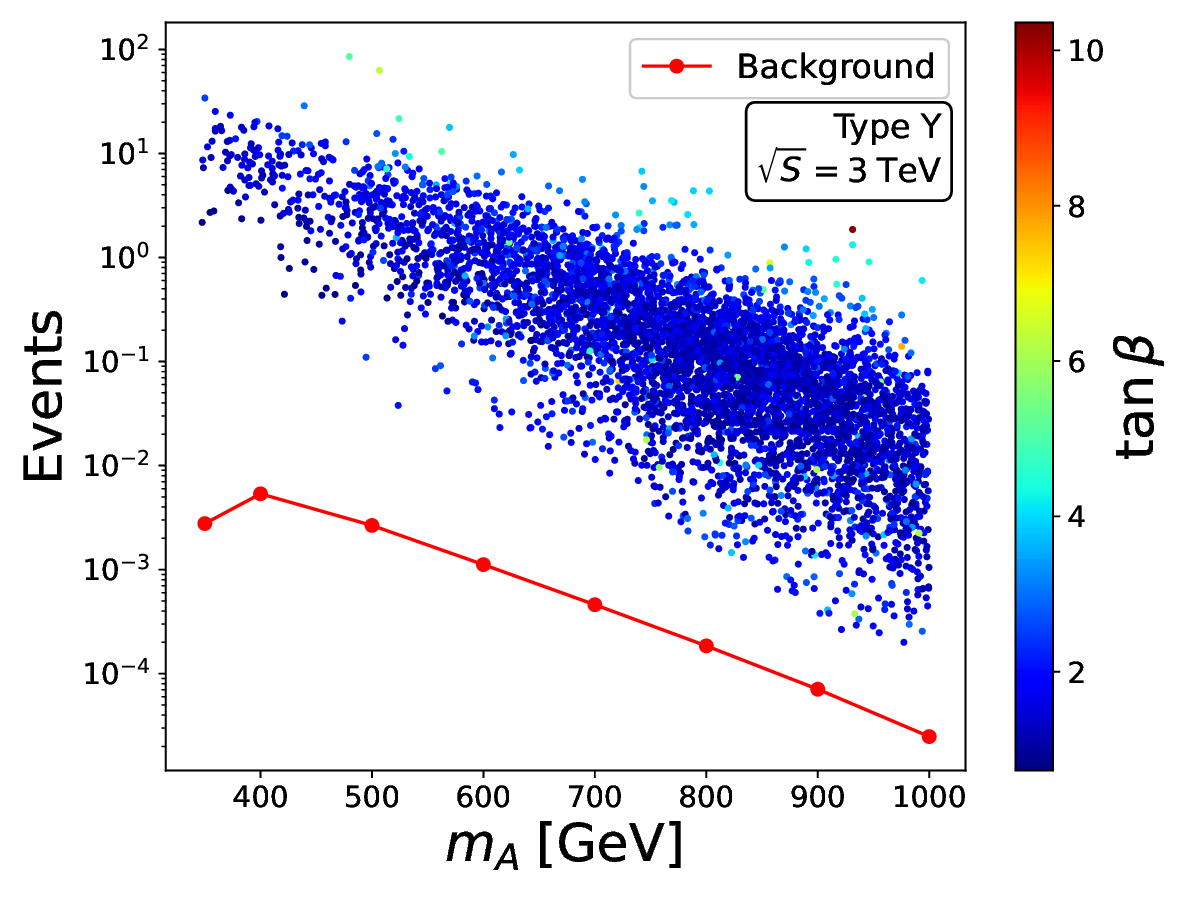}
&
\includegraphics[width=8.5cm,height=6cm]
{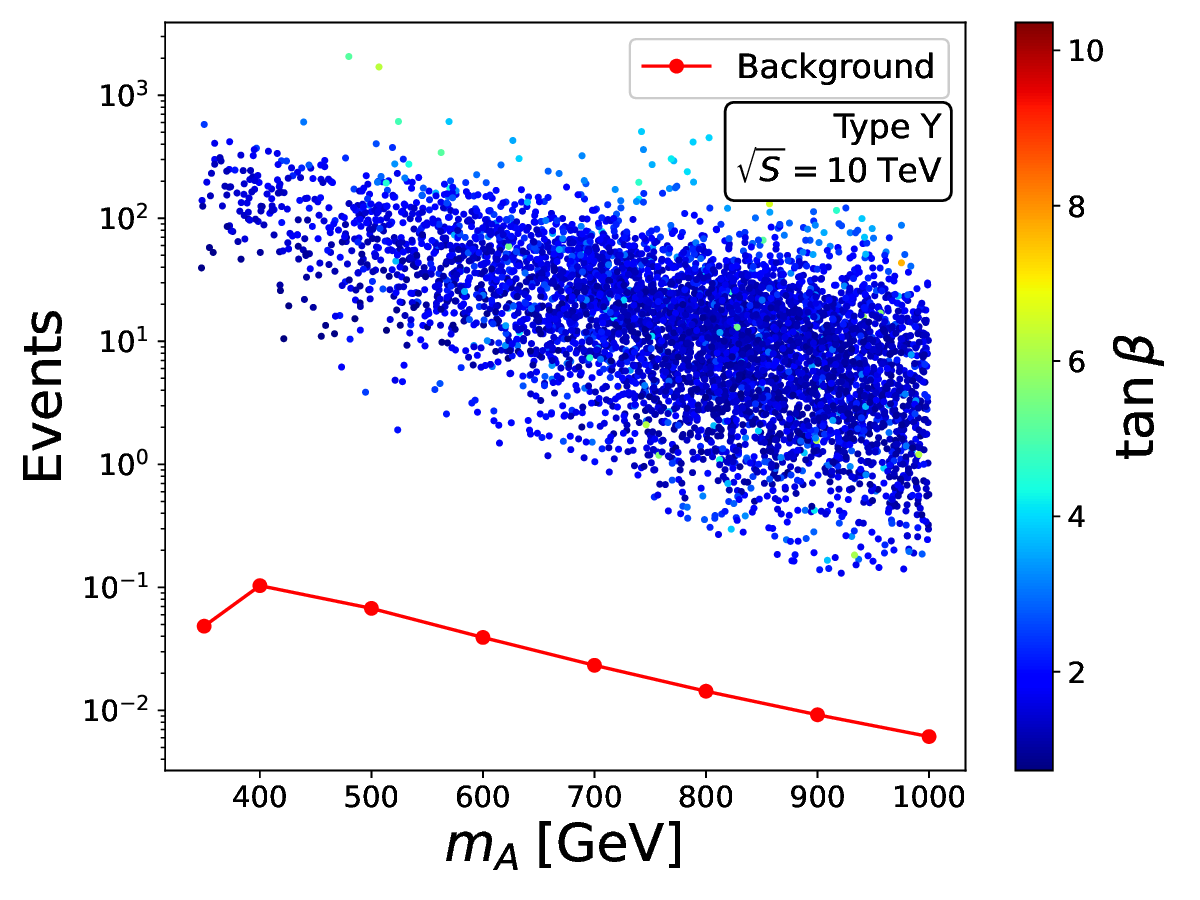}
\end{tabular}
\caption{\label{events}
The event distributions at $\sqrt{s}=3$ TeV 
with an integrated luminosity of 
$\mathcal{L}=3000~\text{fb}^{-1}$ 
and at $\sqrt{s}=10$ TeV 
with $\mathcal{L}=10000~\text{fb}^{-1}$. 
The results for the Type-X THDM are shown 
in the upper two panels, while those for the 
Type-Y THDM are displayed in the lower 
two panels.}
\end{figure}

The branching ratios for the decays $A \to t\bar{t}$
(left panel) and $A \to b\bar{b}$ (right panel) are
shown in the parameter space of $m_A$ and $\tan\beta$.
The results for the Type-X THDM are presented in the
upper panels, while the corresponding decay rates for
the Type-Y THDM are shown in the lower panels. In all
cases, the decay $A \to t\bar{t}$ is dominant compared
to $A \to b\bar{b}$. We also find that the branching
ratios for $A \to b\bar{b}$ in the Type-Y THDM are
larger by more than two orders of magnitude than those
in the Type-X THDM. This behavior can be understood
from the different Yukawa coupling structures in the
two models.
\begin{figure}[H]
\centering
\begin{tabular}{cc}
\includegraphics[width=8.5cm,height=6cm]
{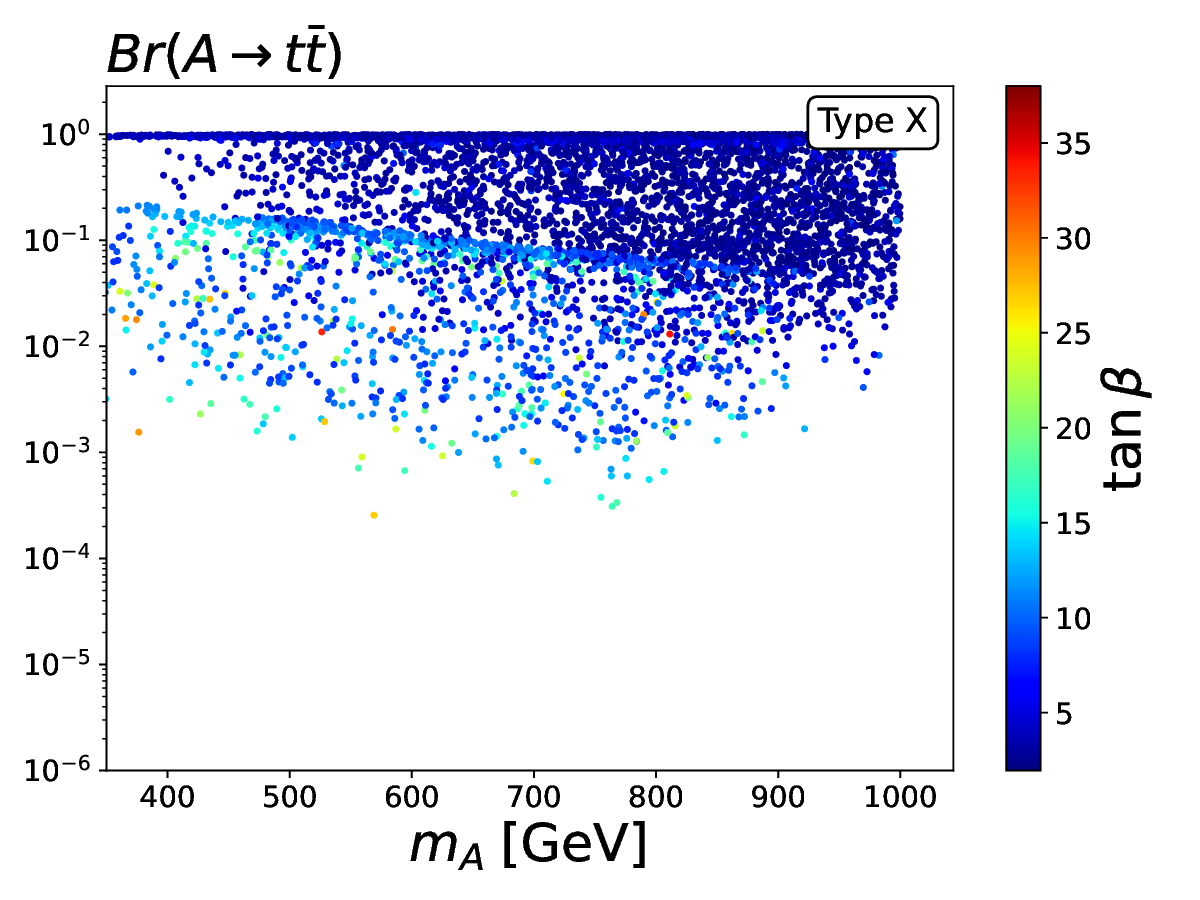}
& 
\includegraphics[width=8.5cm,height=6cm]
{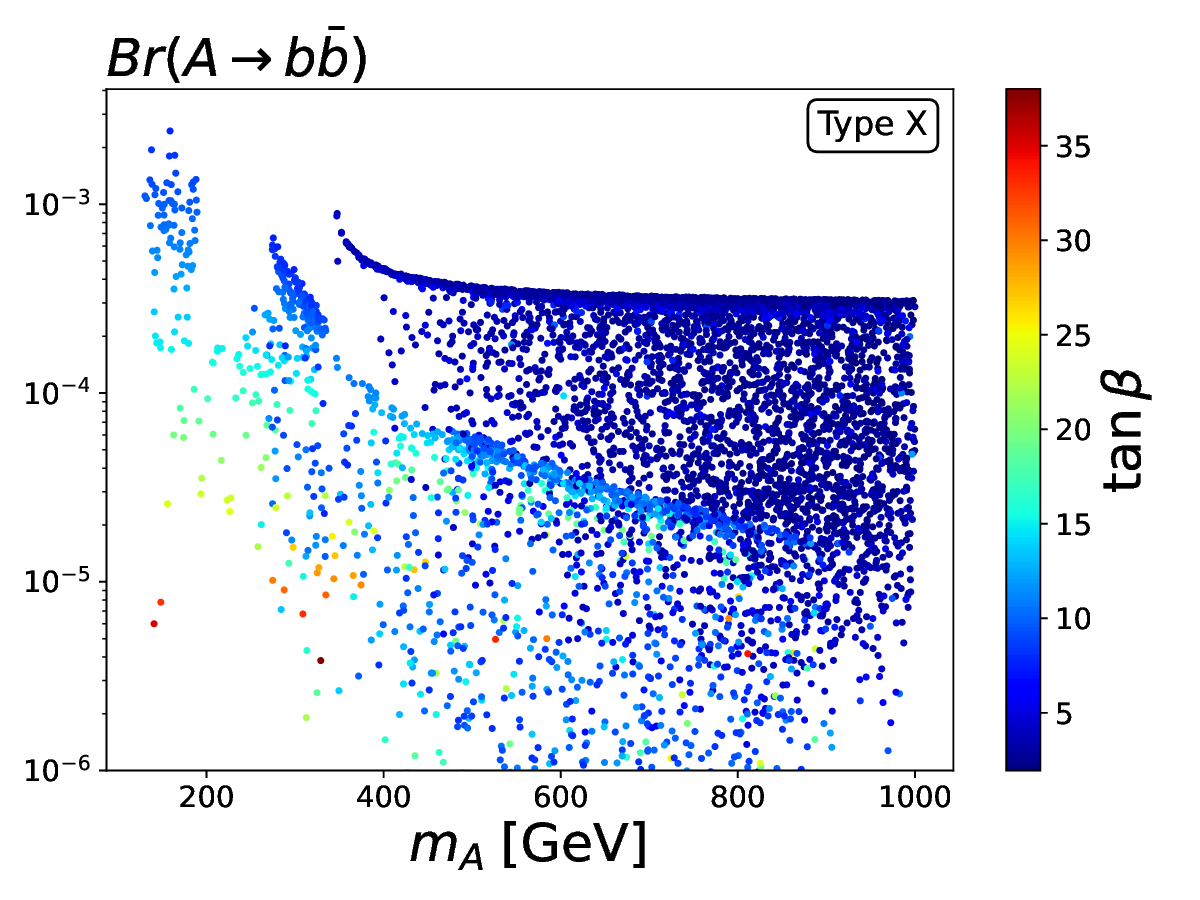}
\\
\includegraphics[width=8.5cm,height=6cm]
{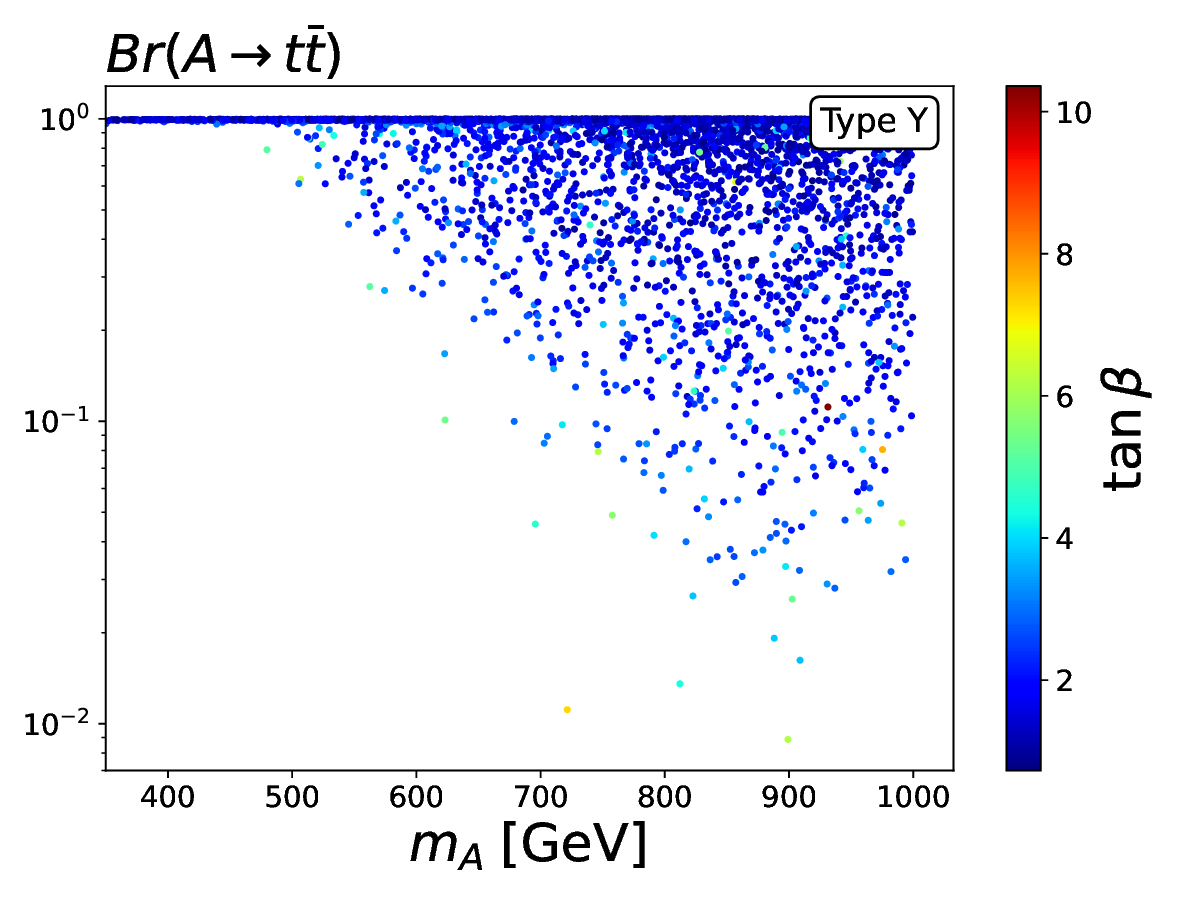}
& 
\includegraphics[width=8.5cm,height=6cm]
{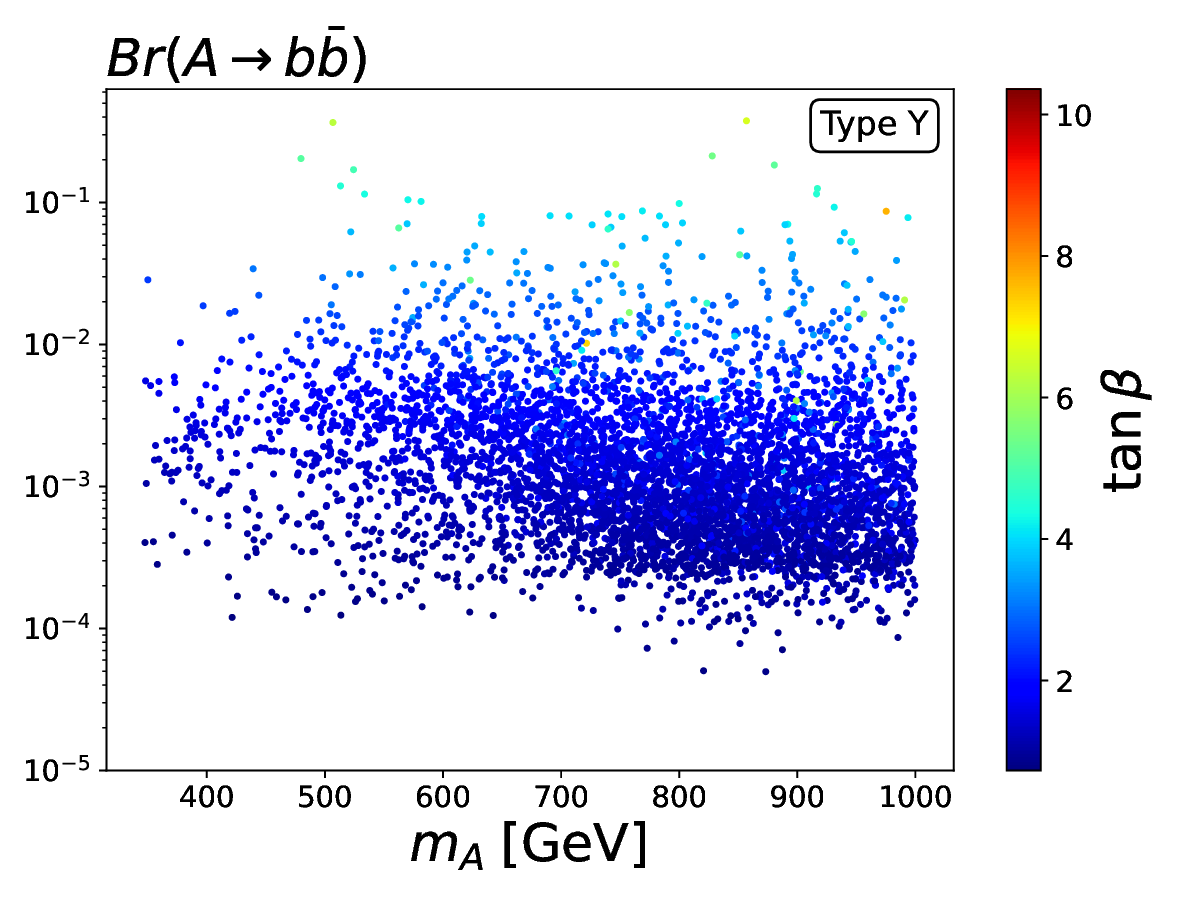}
\end{tabular}
\caption{\label{branching}
The branching ratios for the 
decays $A \to t\bar{t}$ (left panel) 
and $A \to b\bar{b}$ (right panel) 
are shown in the parameter space of 
$m_A$ and $\tan\beta$. The results for 
the Type-X THDM are presented in 
the upper panels, while the corresponding 
decay rates for the Type-Y THDM 
are shown in the lower panels.}
\end{figure}
We turn our attention to the generation 
of events for $\mu^- \mu^+ \to W^\pm W^\mp 
\to A(p_3)A(p_4) \to t\bar{t} b\bar{b}$, 
which are scanned over the parameter space 
of the Type-X and Type-Y THDMs. 
In Fig.~\ref{events}, we present the event 
distributions at $\sqrt{s}=3$ TeV with an 
integrated luminosity of 
$\mathcal{L}=3000~\text{fb}^{-1}$ and at 
$\sqrt{s}=10$ TeV with $\mathcal{L}=10000~\text{fb}^{-1}$. 
The results for the Type-X THDM are shown in the upper two panels, while those for the Type-Y THDM are displayed in the lower two panels. The events are plotted in the parameter space of CP-odd Higgs mass $m_A$ and the mixing 
angle $\tan\beta$. In these plots, the SM background 
events are shown as red dotted lines. For the SM 
background cross-sections, we consider the corresponding processes $\mu^- \mu^+ \to W^\pm W^\mp \to t\bar{t}
b\bar{b}$. The SM background cross sections are evaluated after applying the appropriate kinematic cuts, as described below. To identify the top- and bottom-quark pairs originating from the CP-odd Higgs boson, we apply invariant-mass cuts on the final-state particles,
$\left| m_{t\bar{t}} - m_A \right| 
\leq 20~\text{GeV},
~\left|m_{b\bar{b}} - m_A \right|
\leq 20~\text{GeV}$. 
These cuts effectively reduce 
the SM background, since the SM 
cross sections are calculated far 
from the dominant resonance peaks
associated with $Z$-boson 
mediating. Furthermore, 
the kinematic cuts 
are applied the transverse momentum 
$p^{b/\bar{b}}_{T}\geq 20$ GeV, 
the rapidity $|\eta^{b/\bar{b}}|
\leq 2.4$ for bottom-quarks 
($\bar{b}$-quarks). 

\begin{figure}[ht]
\centering
\includegraphics[width=14cm,height=7cm]
{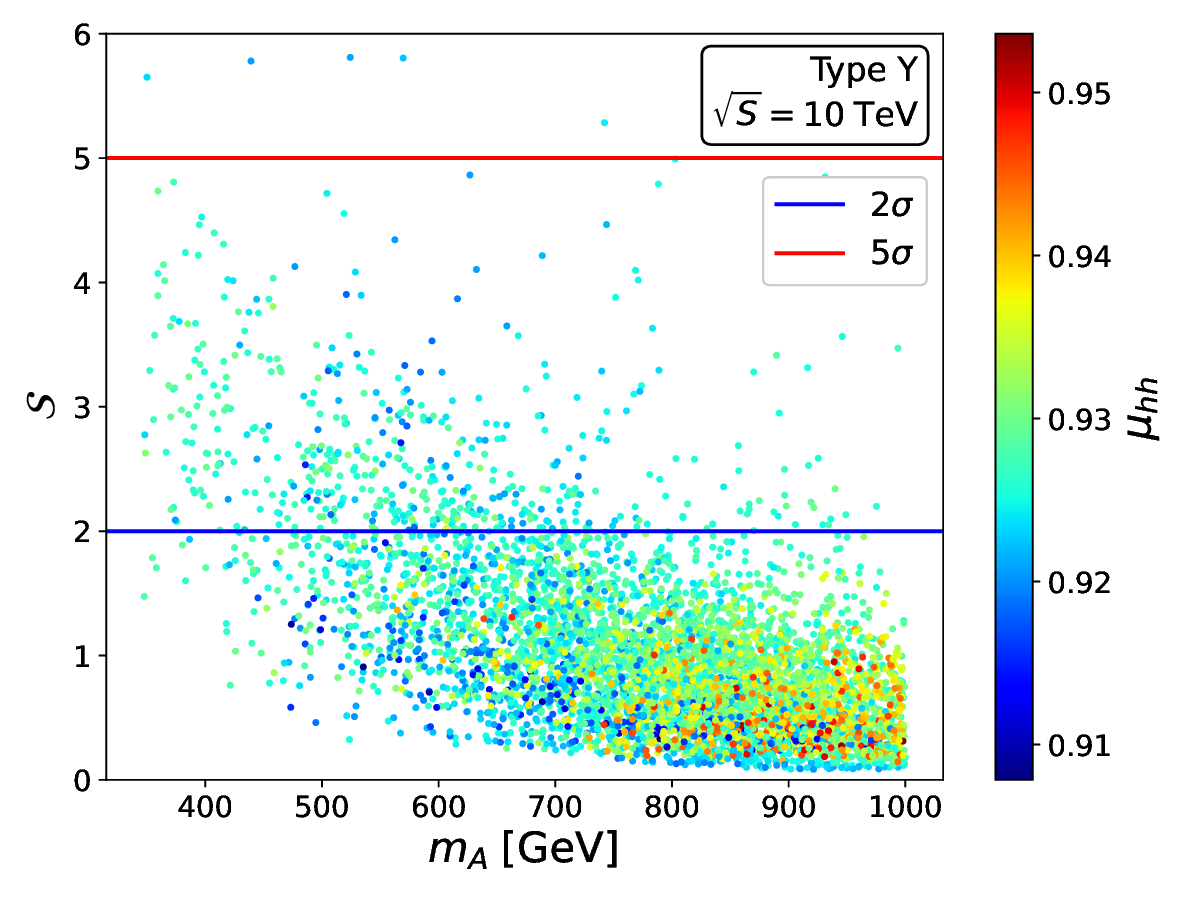}
\caption{\label{SignificanceY}
The statistical significances of 
the signal process $\mu^- \mu^+ 
\to W^\pm W^\mp \to AA \to t\bar{t} b\bar{b}
\to \ell \bar{\ell}\, b\bar{b}\, b\bar{b}$
including missing energy are scanned over the valid
parameter space in correlation with the enhancement
factor $\mu_{hh}$ for the Type-Y THDM at
$\sqrt{s} = 10$~TeV.}
\end{figure}

In all cases, the pair production of CP-odd 
Higgs bosons in the Type-Y THDM at $\sqrt{s}=10$ TeV
multi-TeV muon collider is a promising channel
that can be tested at future colliders. 
In this case, we observe that the 
Standard Model background is rather small 
and it can therefore be neglected.
Moreover, we consider the top-quark decay
into a charged lepton, its associated neutrino, 
and a bottom quark, for which the total
branching fraction is $0.332$ (taken from 
\cite{ParticleDataGroup:2024cfk}). 
The final state is $\ell \bar{\ell}\, b\bar{b}\, b\bar{b}$ 
accompanied by missing energy.
The statistical significance for the process
$\mu^- \mu^+ \to W^\pm W^\mp \to AA \to t\bar{t} b\bar{b} \to \ell \bar{\ell}\, b\bar{b}\, b\bar{b}$
including missing energy is estimated as
$\mathcal{S} = \sqrt{
\mathcal{L} \cdot \sigma \times
\left[\mathrm{Br}(t \to \ell \nu_\ell b)\right]^2 }$.
In Fig.~\ref{SignificanceY}, the 
significances are scanned over the valid parameter
space in correlation with the enhancement factor $\mu_{hh}$. 
We find that the signal can be observed with a significance exceeding $2\sigma$ 
at several parameter points within the viable 
parameter space of the Type-Y THDM. 
However, almost all viable parameter 
points lie below the $5\sigma$ significance 
level.
\section{Conclusions} 
We present the first results for full one-loop
electroweak radiative corrections to
$\mu^- \mu^+ \to W^\pm W^\mp \to hh$ in 
the Standard Model in this work. 
We then evaluate neutral scalar pair 
production through vector
boson fusion at multi--TeV muon colliders 
in the THDM. 
In the phenomenological
analysis, the enhancement factors for SM-like Higgs
pair production in the THDM with respect to the
corresponding ones in the SM are scanned over the
parameter space within the viable regions of the
Type-X and Type-Y THDMs. Our findings show that the
ratio can reach a factor of $3$ in several regions
within the valid parameter space of the Type-X THDM,
whereas it lies in the range $0.91$ to $0.96$ over the
entire parameter space of the Type-Y THDM. Finally, we
evaluate CP-odd Higgs pair production through vector
$W$-boson fusion at multi--TeV muon colliders.
In the Type-Y case at $\sqrt{s} = 10$~TeV
with an integrated luminosity of
$\mathcal{L} = 10000~\text{fb}^{-1}$,
CP-odd Higgs pair production
in the $t\bar{t}b\bar{b}$ final state,
with subsequent top-quark decays into leptons
and bottom quarks taken into account,
can be tested with a statistical significance
exceeding the $2\sigma$ level at several
viable parameter points.
This provides a promising possibility to probe the
scalar Higgs sector and deepen our understanding of
electroweak symmetry-breaking dynamics in particle
physics.
\\

\noindent
{\bf Acknowledgment:}~
This research is funded by Vietnam
National Foundation for Science and
Technology Development (NAFOSTED) under
the grant number $103.01$-$2023.16$.
\section*{Appendix A: Numerical Consistency Checks}
In this paper, we use the input parameters
from Particle Data Gruop \cite{ParticleDataGroup:2024cfk}. 
We are working 
in $G_{\mu}$-scheme where the Fine-Structure Constant
is calculated as 
\begin{eqnarray}
\alpha = \sqrt{2} G_F m_W^2
\left(1-\frac{m_W^2}{m_Z^2} 
\right)/\pi
\end{eqnarray}
with $G_F=1.1663785\cdot 10^{-5}$ GeV$^{-2}$. 

In this Appendix, we present numerical consistency
checks for the one-loop radiative corrections to
the process $\mu^- \mu^+ \to W^\pm W^\mp \to hh$ in the SM.
The {\tt GRACE-Loop} system implements non-linear gauge (NLG)
fixing terms in the Lagrangian. As a result, the total cross
section must be independent of the NLG parameters,
namely $\alpha$, $\beta$, $\cdots$, and $\kappa$.
In the context of one-loop renormalization, the total cross
section is also free of ultraviolet (UV) divergences
($C_{UV}$). Due to the presence of virtual photon exchange
in the loop, the corresponding Feynman diagrams contain
infrared divergences. By including soft-photon radiation,
the cross sections become independent of $\lambda$
(the photon mass). However, the inclusion of soft-photon
radiation makes the cross section dependent on the photon
energy cutoff parameter, denoted by $k_c$. At this stage,
the dependence on $k_c$ is canceled by the contribution
from hard-photon radiation. In the final result, the fully
corrected cross sections are free of all the above-mentioned
parameters.

In Table~1, we present numerical checks of the NLG parameters
$\alpha$, $\beta$, $\cdots$, and $\kappa$, as well as
$C_{UV}$ and $\lambda$. The results show that the cross
sections exhibit good stability when these parameters are
varied over wide ranges. It is noted that we apply
$p_T^{h} \geq 20$~GeV and pseudorapidity
$|\eta_{h}| < 2.4$. The tests are performed at a
center-of-mass energy of $3$~TeV at a future multi--TeV
muon collider.
\begin{table}[H]
\begin{center}
\begin{tabular}
{llll}
\hline\hline
$(\{\alpha, \beta, \cdots, \kappa\}, 
 C_{UV}, \lambda)$
& 
$
\sigma_{\mu^- \mu^+ 
\to W^\pm W^\mp \to hh}^{V} 
$
& $
\sigma_{\mu^- \mu^+ 
\to W^\pm W^\mp \to hh}^{S} 
$
&
$\sigma_{\mu^- \mu^+ 
\to W^\pm W^\mp \to hh}^{V+S}$ 
\\ \hline
$(
\{0,0,0,0,0\}, 0, 10^{-13}
)
$
&
$-2.97(2)\cdot 10^{-4}$ 
&
$2.61(7)\cdot 10^{-4}$ 
&
$-0.35(6) \cdot 10^{-4}$
\\
\hline
$(
\{1,2,3,4,5\}, 10^2, 10^{-15}
)
$
&
$-3.50(5)\cdot 10^{-4}$ 
&
$3.15(0)  \cdot 10^{-4}$
&
$-0.35(5) \cdot 10^{-4}$
\\
\hline
$(
\{10,20,30,40,50\}, 10^3, 10^{-17}
)
$
&
$-4.04(0) \cdot 10^{-4}$ 
&
$3.68(4)  \cdot 10^{-4}$ 
&
$-0.35(6) \cdot 10^{-4}$
\\
\hline
\hline
\end{tabular}
\caption{
\label{NLG-CUV-Lam}
Numerical consistency checks for the one-loop radiative
corrections to the process
$\mu^- \mu^+ \to W^\pm W^\mp \to hh$ in the SM are presented.
In this table, we vary $(\alpha, \beta, \cdots, \kappa)$,
$C_{UV}$, and $\lambda$ in the first column. The second
(third) column corresponds to the virtual one-loop
corrections (soft-photon radiation) in pb, while the last
column shows the sum of these contributions in pb. We note
that $\sigma_{\mu^- \mu^+ \to W^\pm W^\mp \to hh}^{T}
= 9.64(3) \times 10^{-4}$~pb.
}
\end{center}
\end{table}
In Table~\ref{kc-stability test}, tests of the
$k_c$ stability of the results are shown. In the
first column, we vary $k_c$ from $10^{-3}$~GeV
to $10^{-5}$~GeV. The soft-photon contributions
are presented in the second column, while the
hard-photon contributions are shown in the third
column. The last column shows the sum of these
cross sections. All cross sections are given in
pb. We fix $(0,0,0,0,0,0,10^{-15})$ for this test.
We find that the results exhibit good stability
under variations of the photon energy cutoff
parameter.
\begin{table}[H]
\begin{center}
\begin{tabular}
{c@{\hspace{1cm}}l@{\hspace{1cm}}l
@{\hspace{1cm}}l}
\hline\hline
$k_c$ [GeV]
& 
$
\sigma_{\mu^+ \mu^- \to W^\pm W^\mp \to hh}^{S} 
$
& 
$
\sigma_{\mu^+ \mu^- \to W^\pm W^\mp \to hh}^{H} 
$
&
$\sigma_{\mu^+ \mu^- \to W^\pm W^\mp \to hh}^{S+H}$ 
\\ \hline
$ 10^{-3} $
&
$3.15(0) \cdot 10^{-4}$
&
$1.315(2) \cdot 10^{-4}$ 
&
$4.46(5) \cdot 10^{-4}$ 
\\
\hline
$ 10^{-4} $
&
$2.88(3)\cdot 10^{-4}$
&
$1.581(8)\cdot 10^{-4}$
&
$4.46(5)\cdot 10^{-4}$
\\
\hline
$ 10^{-5}$
&
$2.61(7) \cdot 10^{-4}$
&
$1.84(8) \cdot 10^{-4}$
&
$4.46(5) \cdot 10^{-4}$
\\
\hline
\hline
\end{tabular}
\caption{\label{kc-stability test}
Test of the $k_c$ stability of the results. 
In the first column, we change $k_c$ from $10^{-3}$ GeV 
to $10^{-5}$ GeV. The soft-photon results are presented 
in the second column, while the hard contributions 
are shown in the third column. 
The last column shows 
the sum of these cross sections. 
All cross sections are shown in pb. 
We fix $(0,0,0,0,0,0,10^{-15})$ 
for the test.}
\end{center}
\end{table}
\section*{Appendix B: Comparison with 
previous references}
The tree-level process $\mu^- \mu^+ \to W^{\pm} W^\mp \to hh$ in the SM has been studied in many earlier works. We cross-check the tree-level cross section computed in this work with the results reported in Ref.~\cite{Ruiz:2021tdt}. In Table~\ref{EVA}, the first column shows the center-of-mass energies ranging from $4$ TeV to $30$ TeV. The second column presents our results obtained in this work, while the third column lists the results from Ref.~\cite{Ruiz:2021tdt}. The data indicates that our results are in good agreement with those of Ref.~\cite{Ruiz:2021tdt}. 
The last column shows the difference between the two results,
expressed in percent. We find that the difference between the
two results is only a few percent. This is due to the scale
dependence of the results, which is also at the level of a few
percent, as pointed out in Ref.~\cite{Ruiz:2021tdt}.
No cuts are applied to the final-state Higgs bosons for the results
shown in the first row, whereas the results in the second row include
the additional requirement that the invariant mass of the two
final-state Higgs bosons be greater than $1~\text{TeV}$.
\begin{table}[H]
\begin{center}
\begin{tabular}
{c@{\hspace{1cm}}c@{\hspace{1cm}}c
@{\hspace{1cm}}c}
\hline\hline
$\sqrt{s}$ [TeV]
& 
This work 
& 
Ref.~\cite{Ruiz:2021tdt}
& 
$\delta$[\%]
\\
\hline
\hline
$4$
&
$
1.9374
\pm 0.0008
$
&
$2.03$
& 
$4.6\%$
\\
&
$
0.341
\pm 
0.002
$
&
$0.364$
&
$
6.3\%
$
\\
$14$
&
$5.830 
\pm 0.003$
&
$6.01$
&
$3.0\%$
\\
&
$
2.391 
\pm 0.001
$
&
$2.44$
&
$
2.0\%
$
\\
$30$
&
$9.390
\pm 0.006$
&
$
9.48
$
&
$
0.95\%
$
\\
&
$
4.784
\pm 
0.003
$
&
$
4.63
$
&
$
3.1\%
$
\\
\hline
\hline
\end{tabular}
\caption{\label{EVA}
Tree-level process $\mu^- \mu^+ 
\to W^{\pm} W^\mp \to hh$ 
in this work and in 
Ref.~\cite{Ruiz:2021tdt}.
The cross sections are shown in fb.
The results in the second row are 
the same as those in the first row,
but with the additional requirement 
that the invariant mass of the
two final-state Higgs bosons be 
greater than $1~\text{TeV}$.
}
\end{center}
\end{table}
\end{document}